\documentclass[12pt,american]{article}
\usepackage{mathpazo}

\usepackage[latin9]{inputenc}
\usepackage{color}
\usepackage{amsmath}
\usepackage{amssymb}
\usepackage{graphicx}

\makeatletter

\usepackage{setspace} % load setspace before footmisc
\usepackage{footmisc}
\usepackage{latexsym}
\usepackage{amsfonts}

\usepackage{filecontents}
\usepackage{natbib}
\usepackage[citecolor=blue, colorlinks=true, linkcolor=blue]{hyperref}

\usepackage{graphics}\usepackage{amsthm}

\usepackage{mathrsfs}\usepackage{bbm}
\usepackage{subfigure}\usepackage{psfrag}%extra spacing
\usepackage[left=1in, right=1in, top=1in, bottom=1in]{geometry} 
\usepackage{setspace}

\newcommand{\ds}{\displaystyle}

\theoremstyle{plain}

\newtheorem{theorem}{Theorem}%[section]

\newtheorem{lemma}{Lemma}%[section]
\newtheorem{proposition}{Proposition}%[section]

\theoremstyle{definition}
\newtheorem{definition}{Definition}%[section]

\makeatother

\usepackage{authblk}

\title{Transitional Dynamics of the Saving Rate and\\ Economic Growth\thanks{This is a substantially revised version of an earlier paper ``Effects of Income Growth on Domestic Saving Rates: The Role of Poverty and Borrowing Constraints.'' We acknowledge financial support of Singapore Ministry of Education Academic Research Fund Tier 1 (R-122-000-112-112), the Lee Kuan Yew School of Public Policy, National University of Singapore, City University of New York PSC-CUNY Research Award (60124-00 48), Korea University (K2009811) and BrainKorea21 Plus (K1327408 and T192201). We thank Hiro Kasahara, Aditya Goenka, Liugang Sheng and Chong Kee Yip for their valuable comments. Corresponding author: Tomoo Kikuchi. Email: \texttt{tomookikuchi@waseda.jp}.
}}

\author[a]{Markus Brueckner}
\author[b]{Tomoo Kikuchi}
\author[c]{George Vachadze}

\affil[a]{\small Research School of Economics, Australian National University}
\affil[b]{\small Graduate School of Asia-Pacific Studies , Waseda University}
\affil[c]{\small Department of Economics, College of Staten Island, City University of New York}

\begin{document}

\maketitle

\begin{abstract} 
\noindent We estimate the relationship between GDP per capita growth and the growth rate of the national saving rate using a panel of 130 countries over the period 1960-2017. We find that GDP per capita growth increases (decreases) the growth rate of the national saving rate in poor countries (rich countries), and  a higher credit-to-GDP ratio decreases the national saving rate as well as the income elasticity of the national saving rate. We develop a model with a credit constraint to explain the growth-saving relationship by the saving behavior of entrepreneurs at both the intensive and extensive margins. We further present supporting evidence for our theoretical findings by utilizing cross-country time series data of the number of new businesses registered and the corporate saving rate.   

\vspace{1ex}


\noindent \textbf{JEL\ Classification:} D9; E2; O1

\end{abstract}

\clearpage

\section{Introduction}

There is wide evidence that a country's saving rate first rises then falls as the economy grows. In other words, the saving rate exhibits a hump-shape over time \citep[see][]{antras2001transitional}. To explain this stylized fact the literature has modified neoclassical models to include non-homothetic preferences, adjustment costs and structural changes among others \citep[e.g.][]{christiano1989understanding, king1993transitional, laitner2000structural, chen2006japanese}. More recently, \cite{buera2013financial} modified a neoclassical model to include heterogeneous agents and credit constraints to show that a country's saving rate follows a hump-shaped transitional dynamics after a reform that eliminates taxes and subsidies that distort the allocation of resources. This paper contributes to the literature as follows. First, we verify the hump-shaped saving rate of countries and the role of credit constraints 
by a large panel data analysis. Second, we build a theoretical model with a credit constraint to explain how the number of entrepreneurs and their saving behavior induce the hump-shaped relationship between the national saving rate and GDP per capita. Lastly, we provide supporting evidence for this mechanism by utilizing cross-country time series data of the number of new businesses registered and the corporate saving rate.

We start our analysis by estimating the relationship between per annum GDP per capita growth and the per annum growth rate of the national saving rate using panel data that cover 130 countries over the period 1960-2006.\footnote{We provide estimates of the growth-saving relationship over the post-global financial crisis period 2007-2017 in Appendix \ref{sec:post-global-crisis}.} The panel model estimates show that GDP per capita growth significantly increases the growth rate of the national saving rate in poor countries. However, the opposite is the case in rich countries. The estimated effects are quantitatively large. To illustrate their size, consider a low-income country with a GDP per capita (PPP based) of USD 1,000. For this country, our estimates suggest that a 1 percentage point increase in  GDP per capita growth increases the growth rate of the national saving rate by about 4 percentage points. On the other hand, for a high-income country with GDP per capita of USD 50,000 a  1 percentage point increase in GDP per capita growth decreases the growth rate of the national saving rate by about 2 percentage points. 

Furthermore, we find that the credit-to-GDP ratio (the GDP share of domestic credit to the private sector) significantly decreases the national saving rate: a 1 percentage point decrease in the growth rate of the credit-to-GDP ratio increases the growth rate of the national saving rate by around 0.2 percentage points. In addition, the effect of GDP per capita growth on the growth rate of the national saving rate decreases when the credit-to-GDP ratio is higher. So much so, that in countries with a low credit-to-GDP ratio GDP per capita growth increases the growth rate of the national saving rate while in countries with a high credit-to-GDP ratio the opposite is the case.

To explain our empirical findings we develop a model that introduces intertemporal saving decisions into the overlapping generations model by \cite{matsuyama2004financial}.\footnote{In a standard two-period overlapping generation model, output in every period is produced by combining capital provided by the old and labor provided by the young. The young then receive the wage---the wealth of the young---and decide how much to save. This is in contrast to a standard growth model, in which  the wage, saving and output are determined simultaneously by a representative agent, who maximizes utility over an infinite horizon.} The credit constraint in the Matsuyama model gives rise to a positive rent for entrepreneurial activities but not all agents can become entrepreneurs in the presence of indivisible investment. The ex-ante identical young are endogenously divided into {\em entrepreneurs} and {\em investors}. In the Matsuyama model and its extensions \citep[e.g.][]{kikuchi2009endogenous, kikuchi2015financial, kikuchi2018volatile}, however, agents consume only when old and saving is inelastic (the young save their entire wage), i.e., saving is independent of the credit constraint and the wealth.\footnote{This means the young in the Matsuyama model are credit-rationed.} 

Once we allow agents to consume in both periods in the Matsuyama model, both the credit constraint and the wealth affect saving of the young. This requires us to derive an equilibrium that is compatible with preferences of all agents, i.e. both entrepreneurs and investors. The entrepreneurial rent now gives entrepreneurs incentives to save more than investors to overcome the credit constraint and fund their investment projects.\footnote{\cite{kikuchi2018minimum} and \cite{hillebrand2018bubbles} introduce intertemporal decision into the Matsuyama model too. However, \cite{kikuchi2018minimum} only  considers a symmetric equilibrium in which everyone chooses the same level of saving and \cite{hillebrand2018bubbles} considers a model, in which there are a fixed fraction of different types of agents. To the best of our knowledge, our model is the only extension of the Matsuyama model, in which ex-ante identical agents make different saving decisions. } 
Hence, tighter credit constraints increase the saving rate as well as the income elasticity of the national saving rate. When the wealth increases, more agents can become entrepreneurs as they require less external finance. On the other hand, when the wealth increases, entrepreneurs save less. When the wealth is low, the extensive margin dominates the intensive margin; the opposite is true when the wealth is high. Because of this interplay of the intensive and extensive margins GDP per capita 
growth increases the national saving rate in poor countries while the opposite is true in rich countries.

To provide empirical support for the mechanism of the theoretical model, we use cross-country time series data of the number of new businesses registered and the corporate saving rate. The number of new businesses registered is a proxy for the extensive margin in the theoretical model. The corporate saving rate is a proxy for the intensive margin. Our panel model estimates show a significant positive effect of GDP per capita growth on the growth rate of new businesses registered, and that this effect is significantly larger for poorer countries. Moreover, GDP per capita growth has a significant negative effect on the growth rate of the corporate saving rate; and more so in richer countries. 

These empirical results, which are consistent with the mechanism of the theoretical model, suggest that in poor countries the extensive margin is dominant---saving  at the intensive margin plays only a minor role. It is the dominance of  the extensive margin that gives rise to a significant positive effect of GDP per capita growth on the national saving rate in poor countries. On the other hand, in rich countries the intensive margin dominates the extensive margin. That is why in rich countries there is a significant negative effect of GDP per capita growth on the growth rate of the national saving rate: in countries where GDP per capita is already high, entrepreneurs rely less on external funds so that when GDP per capita increases even more, there is a large decrease of the corporate saving rate  and only a small increase in the number of new businesses registered.

Our analysis also speaks to the literature on the role of financial development. One view in this literature is that credit constraints inhibit capital accumulation by preventing a more efficient allocation of credit to investment.\footnote{See \cite{levine1997financial} for a comprehensive survey of both the theoretical and empirical literature. The literature has identified different channels, through which credit constraints may adversely affect output per capita. \cite{galor1993income}, for example, show that credit constraints can create persistence in initial wealth inequalities by preventing children of poor families from obtaining human capital. Credit constraints can also reduce occupational mobility \citep[e.g.][]{banerjee1993occupational, aghion1995bolton, piketty1997dynamics} and prohibit high ability workers from becoming entrepreneurs \citep[e.g.][]{lloyd2000enterprise, matsuyama2000endogenous}} 
The other view is that the effects of credit constraints on savings rates are positive. In particular, it is well-known from the life-cycle literature that credit constraints may increase saving rates 
\citep[e.g.][]{bewley1986stationary, deaton1991saving, aiyagari1994uninsured, levine2002does}. Closely related to our paper with regard to the effects that credit constraints have on saving are \cite{jappelli1994saving} and \cite{ghatak2001occupational}. 
\cite{jappelli1994saving} examines a three-period overlapping generations model, in which agents work, only when middle-aged, but consume in all
three periods. When the credit constraint is binding, the consumption of the young is sub-optimal; however, a tighter credit constraint raises the saving of the middle-aged. In the presence of this trade-off, the authors show the existence of an optimal level of credit constraints.
In a similar vein, \cite{ghatak2001occupational} analyzes a two-period overlapping generations model, with moral hazard in the labor market and transaction
costs in the credit market. They show that higher transaction costs in the credit market induce the young to work harder. The increase in the work-effort by the poor young allows them to overcome the transaction costs and enjoy entrepreneurial rents when old.

The credit constraint in our model, as in \cite{jappelli1994saving} and \cite{ghatak2001occupational}, has a positive effect on the natinoal saving rate. However, our mechanism  is different. In \cite{jappelli1994saving}, loans to consumers are facilitated between generations and consumers do not change their behavior, even in the presence of a binding credit constraint. In our model, loans are facilitated within one generation, between investors and entrepreneurs, and the young agents have dynamic incentives to save more to become entrepreneurs. The young  supply extra effort to become self-financed entrepreneurs in the \cite{ghatak2001occupational} model while in our model they become entrepreneurs through thrift alone.\footnote{\cite{ghatak2001occupational} refer to the dynamic incentives for the young  to work hard and save in order to become self-financed entrepreneurs as the American Dream effect. The role of the credit constraint for encouraging saving to set up businesses is well documented.  An excellent summary can be found in \cite{ghatak2001occupational}. In a study using US data, \cite{buera2009dynamic} documents that people, who eventually become entrepreneurs, save more than people who expect to remain workers.}

In terms of the mechanism that generates the hump-shaped saving rate, \cite{buera2013financial}'s model is closest to ours. In their model after a reform eliminates distortions in the resource allocation, the economy transits from one steady state to another.\footnote{The reform-triggered transitional dynamics is motivated by the historical accounts of the so-called miracle economies such as China, Japan, Korea, Malaysia, Singapore, Taiwan, and Thailand.} Initially, the productive, high-saving entrepreneurs account for only a small fraction
of the national income, but over time, they start to account for a larger fraction, and the
national saving rate rises. Eventually, the saving rates of the entrepreneurs start to fall as they are less likely to be credit constrained and face the diminishing marginal returns. The resulting hump-shaped transitional dynamics of the saving rate arises because the reform eliminates the distortions faced by ex-ante heterogeneous agents. 
In contrast, in our model there is no misallocation of resources, and 
the credit constraint alone causes ex-ante homogenous agents to behave differently; it is optimal for agents, who choose to become entrepreneurs, to save more.

The rest of the paper is organized as follows. Section \ref{sec:prediction1} presents the baseline regression results. 
Section \ref{sec:prediction23} presents an interaction model to provide further evidence for our findings. Section %\ref{sec:post-global-crisis} presents results for the post-global financial crisis period. 
Section \ref{sec_model} develops the theoretical model.
Section \ref{sec:support-mechnism} presents empirical support for the theoretical mechanism. 
Section \ref{sec_conclusion} concludes. Appendix \ref{sec:cobb-douglas} - \ref{sec:clustering} contain all remaining proofs and some extensions.

\section{The Effects of GDP per capita and the Credit-to-GDP ratio on the Saving Rate \label{sec:prediction1}}

We begin by estimating the average effects that growth in GDP per
capita and the credit-to-GDP ratio (the GDP share of domestic credit to the
private sector) have on the growth rate of the
national saving rate.\footnote{
	Due to non-stationarity of time series of GDP per capita and the savings
	rate, we use first differences of the variables for our regression
	analysis.} The lower is the credit-to-GDP ratio, the tighter is the credit constraint. The econometric model is: 
\begin{equation}
\triangle\ln(s_{it})=\gamma\triangle\ln(y_{it})+\theta\triangle\ln(\lambda_{it})+a_{i}+b_{t}+u_{it}\label{eq:12}
\end{equation}
where $\triangle\ln(s_{it})$ is the year $t-1$ to $t$ change in
the log of the national saving rate; $\triangle\ln(y_{it})$ is the
year $t-1$ to $t$ change in the log of GDP per capita; $\triangle\ln(\lambda_{it})$
is the year $t-1$ to $t$ change in the log of the credit-to-GDP
ratio; $a_{i}$ is a country fixed effect; $b_{t}$ is a year fixed
effect; and $u_{it}$ is an error term that is clustered at the country
level.

The above econometric model can be derived from our theoretical model,
see Section \ref{sec:specification}. In Section \ref{sec:identification}, we discuss
identification issues pertaining to the estimation of the econometric
model. We present and interpret our empirical results in Section \ref{sec:empirical-results}. 

\subsection{Identification Issues\label{sec:identification}}

Potential endogeneity of GDP per capita and the credit-to-GDP ratio
is an important empirical issue in the estimation of (\ref{eq:12}).
Endogeneity biases could arise because a within-country change in
the saving rate  affects GDP per capita and the credit-to-GDP ratio
or because of time-varying omitted variables that affect the savings
rate beyond GDP per capita and the credit-to-GDP ratio. Moreover,
it is well-known that classical measurement error attenuates least
squares estimates towards zero (thus leading to an understatement
of the true causal effect that GDP per capita and the credit-to-GDP
ratio have on the national saving rate). In order to correct for
endogeneity and measurement error bias, we need plausible exogenous
instruments for GDP per capita and the credit-to-GDP ratio. Instrument
validity requires that the instruments should only affect the savings
rate through their effects on the endogenous variables. Because the
estimating equation includes country fixed effects, such instruments
need to be time-varying.

We use year-to-year variations in the international oil price weighted
with countries' average net-export share of oil in GDP as an instrument
for the change in GDP per capita.\footnote{See \cite{bruckner2012estimating,bruckner2012oil}. For an application
	of this IV strategy to US states, see \cite{acemoglu2013income}.} It is important to note that because year-to-year variations in the
international oil price are highly persistent \cite[for evidence on international oil prices' random walk behavior]{hamilton2009understanding,bruckner2012oil},
the instrumental variable estimate of $\gamma$ captures the effect
that a persistent shock to countries' GDP per capita has on the growth
rate of the saving rate. Because the oil price shock instrument is
constructed based on countries' average net export shares (i.e. the
net export shares are time-invariant), the time-series variation comes
exclusively from the variation in the international oil price. By
weighting the variation in the international oil price with countries'
average net export shares of oil in GDP, the instrument takes into
account that how the international oil price affects GDP per capita
growth differs across countries, depending on whether they are net
importers or exporters of oil. We can reasonably assume that the majority
of countries are price takers in the international oil market. In
order to ensure that our estimates are not driven by potentially large
oil exporting or importing countries, where the exogeneity assumption
may be more questionable, we will also present estimates that are
based on a sample, which excludes large oil exporters and importers.

We use a lagged credit-to-GDP ratio as an internal instrument for
the current credit-to-GDP ratio. Using the lagged variable as an instrument
should reduce concerns that our within-country estimate of $\theta$
is inconsistent. Moreover, the first difference specification eliminates
omitted variables bias, arising from time-invariant cross-country
differences in historical and geographical variables that may be affecting
the saving rate and the credit-to-GDP ratio.

\subsection{Empirical Results\label{sec:empirical-results}}

We obtained data for national saving rates and GDP per capita (constant
price PPP based) from the Penn World Table \cite{heston2011penn},
and the GDP share of domestic credit to the
private sector from the World Development Indicators \cite{bank2012world}.
The sample consists of 130 countries over the period 1960-2007. For
a list of countries in the sample, see Table \ref{tab:appendix-1}
in Appendix \ref{sec:appendix-table}.

Table \ref{tab:1} presents our baseline estimates of the average
effect that growth in GDP per capita ($y$) and the credit-to-GDP
ratio ($\lambda$) have on the growth rate of the national savings
rate ($s$). In columns (1)-(4), we present instrumental variables
estimates. For comparison, we show in column (5) estimates from the
\cite{pesaran1995estimating} mean-group (MG) estimator and in column
(6) we show estimates from the least squares (LS) fixed effects estimator.
All regressions control for country and year fixed effects (which
are jointly significant at the 1 percent significance level). 
\begin{table}[t]
	\singlespacing
	\footnotesize
	\begin{center}
		\begin{tabular}{lllllll}
			\hline
			
			\multicolumn{7}{c}{$\Delta\ln(s_{it})$} \\
			
			\hline
			& (1) & (2) & (3) &  (4) & (5) &(6)\\
			
			& IV & IV & IV &  IV & MG &  LS\\

			$\Delta\ln(y_{it})$    & 2.50***    &2.43***&  2.47*** & 2.12***  & 1.82*** & 1.52***\\
			
			& (0.81)   & (0.82) & (0.83) & (0.71)   & (0.20)  & (0.15)\\

			$\Delta\ln(\lambda_{it})$    &    &-0.21**&  -0.22** & -1.23*  & -0.98** & -0.26***\\
			
			&  & (0.09) & (0.10) &  (0.68)  & (0.42)  & (0.10)\\

			$\Delta\ln(s_{it-1})$    &  &    &  0.19*** & 0.20*** &  & \\
			
			&   &  & (0.07) &  (0.07)  &  &\\

			Kleibergen-Paap   &  19.60    & 19.40     & 9.96& 3.70 & . & . \\
			
			F-stat  &      &  & & & & \\
			
			Cragg-Donald F-stat   &  283.67    & 274.86     & 136.37& 14.04 & . & . \\
			
			Endogenous    &   $\Delta\ln(y_{it})$  &   $\Delta\ln(y_{it})$ &  $\Delta\ln(y_{it})$,  $\Delta\ln(s_{it-1})$ & $\Delta\ln(y_{it})$, $\Delta\ln(\lambda_{it})$, &. &.\\
			
			Regressors   &     &  &    & $\Delta\ln(s_{it-1})$& &\\

			Instruments   &   $OPS_{it}$  &   $OPS_{it}$ &  $OPS_{it}$, $\ln(s_{it-2})$ &  $OPS_{it}$, $\ln(\lambda_{it-1})$, &. &.\\
			
			&    &   &  &$\ln(s_{it-2})$  & &\\
			
			Country FE  &  Yes    & Yes   & Yes&Yes &  Yes    & Yes   \\

			Year FE        & Yes  & Yes  & Yes & Yes& Yes  & Yes  \\
			
			Observations               & 3781    & 3781 & 3781 &3781&3781  & 3781 \\
			
			\hline
			
		\end{tabular}
	\end{center}
	
	Note: The dependent variable, $\triangle\ln(s_{it})$, is the change in the log of the saving rate. $\triangle\ln(y_{it})$ is the change in the log of real GDP per capita; $\triangle\ln(\lambda_{it})$ is the change in the log of the credit-to-GDP ratio. The method of estimation in columns (1)-(4) is two-stage least squares; in columns (5) mean-group estimation; column (6) least squares fixed effects. Huber robust standard errors (shown in parentheses) are clustered at the country level. \vspace{1em} \caption{\label{tab:1} Effects of growth in GDP per capita and the credit-to-GDP ratio on the growth rate of  the national saving rate}
\end{table}

The main result from the panel regressions is that GDP per capita growth,
on average, has a significant positive effect on the growth rate of
the national saving rate. The growth rate of the credit-to-GDP ratio
has a significant negative effect on the growth rate of the national
saving rate. Specifically, the IV estimates in column (1) show that
unconditional on the credit-to-GDP ratio, the estimated coefficient
on log GDP per capita is 2.5; its standard error is 0.8. Column (2)
shows that the average effect of growth in GDP per capita on the growth
rate of the national saving rate is not much different when we control
for the growth rate of the credit-to-GDP ratio. In columns (3) and
(4), we document that these results are robust to a dynamic panel specification
and instrumenting the change in the credit-to-GDP ratio with its lag.

Columns (5) and (6) show that the MG and LS estimates of $\theta$
($\gamma$) are also negative (positive) and significantly different
from zero at the conventional significance levels. Quantitatively,
the IV estimates are in absolute size somewhat larger than the MG
and LS estimates. One possible reason for this is a classical measurement
error that attenuates the LS and MG estimates but not the IV estimates.
%
%\textcolor{red}{We provide further sensitivity analysis of our IV estimates in 
%Online Appendix. The main result from these robustness checks is that the estimated coefficient of $\theta$ ($\gamma$) is negative (positive) and significantly different from zero at the conventional significance levels. } 

Our first main empirical finding is thus that the  response of the growth rate of the national saving rate to growth in GDP per capita (the credit-to-GDP
ratio) is positive (negative). This is consistent with previous empirical
literature, that has examined the macroeconomic relationship between
saving and GDP per capita \cite{jappelli1994saving,loayza2000drives}.
The finding is also consistent with our theoretical predictions and
suggests that the majority of countries have a level of GDP per capita
and credit-to-GDP ratios, which are relatively low, so that growth
in GDP per capita has a significant positive effect on the growth
rate of  the national saving rate.

\section{Interaction of GDP per capita Growth with GDP per capita and the
	Credit-to-GDP Ratio\label{sec:prediction23}}

\subsection{Estimation Framework}

In this section we present estimates of a model where the income elasticity
of the national saving rate varies as a function of countries' average
GDP per capita and the credit-to-GDP ratio. 
\begin{multline}\label{eq:interactions}
\triangle\ln(s_{it})=\gamma'\triangle\ln(y_{it})+\delta(\triangle\ln(y_{it})*\lambda_{i})\\[2ex]
+\zeta(\triangle\ln(y_{it})*y_{i})+  \theta'\triangle\ln(\lambda_{it})+a'_{i}+b'_{t}+u'_{it}.
\end{multline}

The above econometric model can be derived from our theoretical model (see Section \ref{sec:specification}). 

We use countries' period average credit-to-GDP ratio, $\lambda_{i}$,
to construct the first interaction term. This allows us to focus on
how long-run cross-country differences in the credit-to-GDP ratio
affect the income elasticity of the national saving rate. Note that we construct
the interaction term as $\triangle\ln y_{it}*\lambda_{i}$. Likewise,
we construct the second interaction term that captures the elasticity
as $\triangle\ln(y_{it})*y_{i}$ where $y_{i}$ is a country's average
GDP per capita. The particular construction of the interaction terms
implies that the coefficient $\gamma'$ captures the predicted elasticity
when $\lambda_{i}$ and $y_{i}$ are zero.

\subsection{Empirical Results}

Table \ref{tab:2} presents estimates from the above interaction model.
We begin by reporting in column (1) estimates from a more parsimonious
version of the interaction model that only has as interaction term
$\triangle\ln(y_{it})*\lambda_{i}$. In this model specification,
the estimate of $\gamma'$ captures the predicted income elasticity
of the national saving rate when $\lambda_{i}=0$. Our estimate of
$\gamma'$ is 6.2 and its standard error is 1.3. Thus, the estimate
of $\gamma'$ is positive and significantly different from zero at
the 1 percent level. The estimate of $\delta$ is around -10.8 (s.e.
3.1), hence negative and significantly different from zero at the
1 percent level. The significant negative $\delta$ indicates that
the income elasticity of the national saving rate significantly increases
as countries' credit-to-GDP ratios are lower. So much so, that at
sample minimum, $\lambda_{i}=0.02$, the predicted elasticity is 5.9
with a standard error of 1.3; at sample maximum, $\lambda_{i}=1.51$,
the predicted elasticity is -10.1 with a standard error of 3.5. Column
(2) shows that this result also holds when we control for the direct
effect of $\triangle\ln(\lambda_{it})$ on $\triangle\ln(s_{it})$,
which continues to be negative and significant at the 5 percent level.

We present in columns (3) and (4) estimates of an interaction model
that has as interaction term only $\triangle\ln(y_{it})*y_{i}$. In
this model specification, the estimate of $\gamma'$ captures the
predicted elasticity when $y_{i}=0$. We find that $\gamma'$ is 4.8
and its standard error is 1.2. Thus, the estimate of $\gamma'$ is
positive and significantly different from zero at the 1 percent level.
The estimate of $\zeta$ is -0.13 (s.e.~0.05), hence negative and
significantly different from zero at the 5 percent level. The significant
negative $\zeta$ indicates that the elasticity is significantly higher
in poor countries. Column (4) shows that this result also holds when
we control for the direct effect of $\triangle\ln(\lambda_{it})$
on $\triangle\ln(s{}_{it})$, which continues to be negative and significantly
different from zero at the 5 percent level.

\begin{table}[t]
	\singlespacing
	%	\scriptsize
	\footnotesize
	
	\begin{center}
		\begin{tabular}{lllllll}
			\hline
			
			\multicolumn{7}{c}{$\Delta\ln(s_{it})$} \\
			
			\hline
			& (1)        & (2)      & (3)      &  (4)    & (5) & (6)\\

			$\Delta\ln(y_{it}) $                              & 6.16***    & 6.08***  & 4.82*** & 4.77***  & 6.81*** &6.74***\\
			
			& (1.33)&(1.33)&(1.22)&(1.22)&(1.27)&(1.25)\\
			
			$\Delta\ln(y_{it})*\lambda_i$                    & -10.82***    & -10.87***  & &  & -9.23*** & -9.25***\\
			
			& (3.09)&(3.05)&&&(3.01)&(2.94)\\
			
			$\Delta\ln(y_{it})*y_i$                     &   &  & -0.13*** & -0.13***  & -0.06** &-0.06**\\
			
			& &&(0.05)&(0.05)&(0.03)&(0.03)\\
			
			$\Delta\ln(\lambda_{it})$                               &     & -0.29**  & & -0.20**  & &-0.27**\\
			
			& &(0.12)&&(0.09)&&(0.12)\\
			
			Kleibergen-Paap                   &  18.82     & 18.85    &12.80 & 12.53  & 14.08            &14.05\\
			F-stat                            &&&&&\\

			Endogenous      &   $\Delta\ln(y_{it})$,  &   $\Delta\ln(y_{it})$,  & $\Delta\ln(y_{it})$, &  $\Delta\ln(y_{it})$,   &  $\Delta\ln(y_{it})$,&  $\Delta\ln(y_{it})$, \\

			Regressors&   $\Delta\ln(y_{it})^*\lambda_i$,  &   $\Delta\ln(y_{it})^*\lambda_i$  &$\Delta\ln(y_{it})^*y_i$ &  $\Delta\ln(y_{it})^*y_i$  &
			$\ln(y_{it})^*\lambda_i$,&  $\ln(y_{it})^*\lambda_i$, \\
			
			&  & & & & $\Delta\ln(y_{it})^*y_i$ & $\Delta\ln(y_{it})^*y_i$\\

			Instruments   &   $OPS_{it}$,  &   $OPS_{it}$,&  $OPS_{it}$,  & $OPS_{it}$, & $OPS_{it}$,  &$OPS_{it}$,   \\
			
			&   $OPS_{it}{}^*\lambda_i$   &   $OPS_{it}{}^*\lambda_i$&   $OPS_{it}{}^*y_i$ &  $OPS_{it}{}^*y_i$  &  $OPS_{it}{}^*\lambda_i$, & $OPS_{it}{}^*\lambda_i$,  \\

			& &&&& $OPS_{it}{}^*y_i$  &  $OPS_{it}{}^*y_i$     \\
			
			Country FE                              &  Yes       & Yes       & Yes    & Yes              & Yes  & Yes              \\
			
			Year FE                                 & Yes        & Yes       & Yes    & Yes              & Yes  & Yes              \\
			
			Observations                                       & 3850       & 3850     & 3850  & 3850             &3850 & 3850 \\
			
			\hline
		\end{tabular}
	\end{center}
	
	Note: The method of estimation is two-stage least squares. Huber robust standard errors (shown in parentheses) are clustered at the country level. The dependent variable is the change in the log of the saving rate. {*}Significantly different from zero at the 10 percent significance level, {*}{*} 5 percent significance level, {*}{*}{*} 1 percent significance level. \vspace{1em}
	\caption{\label{tab:2} Interactions between GDP per capita growth, GDP per capita, and the credit-to-GDP ratio
	}
\end{table}

In columns (5) and (6), we present estimates from an econometric model
that includes both interaction terms, $\triangle\ln(y_{it})*y_{i}$
and $\triangle\ln(y_{it})*\lambda_{i}$. In this model, the estimates
of $\delta$ and $\zeta$ capture conditional interaction effects.
That is, $\zeta$ captures how the elasticity varies across poor and
rich countries when holding $\lambda_{i}$, the average credit-to-GDP
ratio, constant. Likewise, $\delta$ captures how the average credit-to-GDP
ratio affects the income elasticity of the national saving rate,
when holding $y_{i}$, countries' average GDP per capita, constant.
Hence, any effect that being rich or poor has on the elasticity within
a country through the credit-to-GDP ratio, is shut down. Likewise,
any effect that the credit-to-GDP ratio may have on the elasticity
through $y_{i}$, the average GDP per capita level, is also shut down.

\begin{figure}[ht!]
	{
		\center
		{
			\scalebox{.6}{\hspace{6cm}\includegraphics{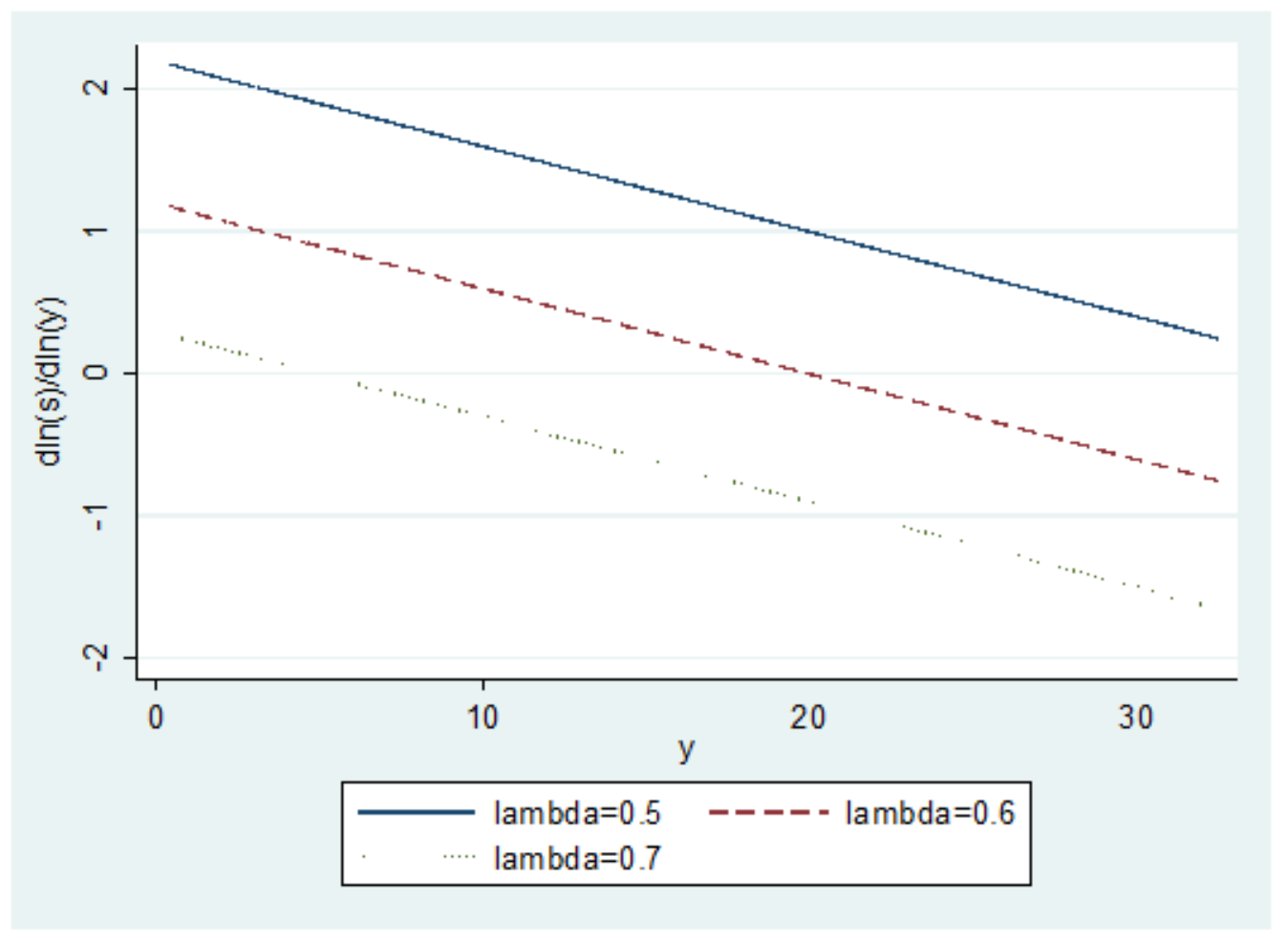}}
		}
	}
	
	\footnotesize
	Note:  The figure is based on the estimates in column (6) of Table \ref{tab:2}.
	
	\caption{The predicted income elasticity of the national saving rate for different levels of average
		GDP per capita and credit-to-GDP ratio. \label{fig:marginal-effects}}
\end{figure}

With the above in mind, we can now interpret the estimates in columns
(5) and (6). The estimate of $\gamma'$ is 6.8, which captures the
predicted income elasticity of the national saving rate when both
$y_{i}=0$ and $\lambda_{i}=0$; its standard error is 1.3. Note that
there is no country in the sample with $y_{i}=0$ and $\lambda_{i}=0$.
It makes more sense to consider countries with very low $\lambda_{i}$
and $y_{i}$, say, those at the bottom 5th percentile ($y_{i}=0.6$
and $\lambda_{i}=0.07$).\footnote{Note that GDP per capita, $y_{i}$, is measured in thousands; see
	also Appendix \ref{sec:robustness} Table \ref{tab:appendix-2} . Hence, $y_{i}=0.6$ refers to a country with
	GDP per capita of USD 600.} For computing the predicted elasticity, we need to consider that
the estimates of $\delta$ and $\zeta$ are -9.2 and -0.06, respectively;
their standard errors are 3.0 and 0.03, respectively. Hence, for a
country at the bottom 5th percentile ($y_{i}=0.6$ and $\lambda_{i}=0.06$),
a 1 percentage point increase in growth of GDP per capita is predicted
to increase the growth rate of the national saving rate by around
6.1 percentage points (s.e. 1.1). This is a large effect. For higher
values of $y_{i}$ and $\lambda_{i}$, the effect is considerably
smaller. For example, at the 50th percentiles ($y_{i}=3.5$ and $\lambda_{i}=0.27$),
the predicted income elasticity of the national saving rate is 4.1,
and at the 75th percentiles ($y_{i}=3.5$ and $\lambda_{i}=0.27$),
it is 1.4. Moreover, the elasticity can be negative for sufficiently
high values of $y_{i}$ (and $\lambda_{i}$). This is shown in Figure
\ref{fig:marginal-effects}, which plots for different values of $\lambda_{i}$
the predicted elasticity over the sample range of $y_{i}$.

Appendix \ref{sec:robustness} shows that our results robust to excluding the top and bottom
1st percentile of the change in the national saving rate; using initial (1970) oil net-export GDP shares to construct the oil price shock instrument; and splitting the sample into the post-1990 and pre-1990 period. Appendix \ref{sec:post-global-crisis} shows that results are similar for the post- and pre-financial crisis period. Appendix \ref{sec:clustering} examines parameter heterogeneity: we report estimates when countries are clustered according to their GDP per capita, geographic location, population size, and sectoral composition.

To summarize: In poor countries growth in GDP per capita has a significant
positive effect  on the growth of the national saving rate. In rich countries and
countries where the credit-to-GDP ratios are high, the opposite is
the case, that is, growth in GDP per capita has a significant negative
effect on the growth rate of the national saving rate.

\section{The Theoretical Model\label{sec_model}}

The economy is inhabited by overlapping generations of agents living for two  periods. Every agent supplies one unit of labor inelastically when young and consumes when both young and old. Successive generations have unit mass and are ex-ante homogeneous by possessing one unit of labor endowment. Production combines the current stock of capital $k_{t}$ supplied by old agents with one unit of labor supplied by young agents.\footnote{``Labor" can be interpreted broadly to include any endowment held by young agents, while ``capital" can be interpreted broadly to include human capital or any other reproducible good used in production.} The resulting output per capita is $y_{t}=f(k_t)$ where $f:R_+ \rightarrow R_+$ is the production function in intensive form. We assume that $f(0)=0$, $f:R_+ \rightarrow R_+$ is twice continuously differentiable on $R_{++}$ and strictly increasing and strictly concave on $R_+$. Factor markets are competitive and rewards on physical capital and labor are determined by the marginal product rule, i.e., $f^{\prime}(k_t)$ is the rate of return on one unit of capital and $w_t=w(k_t):=f(k_t)-k_t f^{\prime}(k_t)$ is the wage. The final commodity produced at time $t$ may be either consumed or invested to produce capital which becomes available in period $t+1$. Capital depreciates fully within a period so that the next period capital stock is equal to the investment.

Following production and distribution of factor payments, the old consume and exit the model, while the young receive the wage--the wealth of the young---and make their \emph{saving} and \emph{investment} decisions. 
When making the saving decision at time $t$, young agents have two options to transfer the current saving into the second period. First, they may become \emph{an investor} by lending the entire saving $s_t^lw_t$ in the competitive credit market for a rate of return $r_{t+1}$. Second, young agents may become \emph{an entrepreneur} by starting an investment project, which requires one unit of the final commodity for investment and returns $R>0$ units of capital in period $t+1$. Later on, we restrict parameter values so that entrepreneurs must always borrow $(1-s_t^bw_t)$ in order to start an investment project. That's why we use the superscript ``$b$'' (borrower) to refer to entrepreneurs and ``$\ell$'' (lenders) to refer to investors. Produced $R$ units of capital can be rented to the final commodity producing firm in exchange for $Rf^{\prime}(k_{t+1})$ units of the final commodity. Hence, the second period consumption of entrepreneurs is $c_{2t+1}=Rf^{\prime}(k_{t+1})-(1-s^b_tw_t)r_{t+1}=(\phi_{t+1}-1+s^b_tw_t)r_{t+1}$ where $\phi_{t+1}:=Rf^{\prime}(k_{t+1})/r_{t+1}$ is the entrepreneurial rent.

We assume that entrepreneurs can pledge only up to a fraction $\lambda \in (0,1)$ of the project revenue for debt repayment. Thus, they can borrow $1-s^b_tw_t$ units of final commodity and start an investment project only if
\begin{equation}\label{eq:bc}
\begin{array}{ccc}
  (1-s^b_{t}w_t)r_{t+1} \leq \lambda R f^{\prime}(k_{t+1}) & \Leftrightarrow & s^b_t \geq \frac{1-\lambda \phi_{t+1}}{w_t}.
\end{array}
\end{equation}
We refer to this as the credit constraint. The parameter $\lambda$ measures the severity of  the credit constraint, with a higher (lower) value corresponding to a looser (tighter) credit constraint. This formulation is a parsimonious way of introducing a credit constraint in a dynamic macroeconomic model. One of the justifications of the credit constraint is the existence of a default cost, which is proportional to the project revenue. In such case, lenders avoid strategic default of borrowers by limiting their debt obligation.\footnote{See footnote 13 in \cite{matsuyama2004financial}.} It is clear from the above inequality that the saving rate of entrepreneurs must be above a threshold, which depends on the severity of the credit constraint ($\lambda$) and on the entrepreneurial rent ($\phi_{t+1}$). The credit constraint binds when the wage is sufficiently low. We will see later that this creates an entrepreneurial rent and motivates entrepreneurs to save more than investors. The higher is the wage, the less do entrepreneurs require external funds, and thus, the lower are their  incentive to save. The difference in saving between entrepreneurs and investors completely disappears as the wage exceeds a threshold.

\subsection{Optimal Behavior\label{sec_optimal}}

Young agents maximize the following lifetime utility
\begin{equation}\label{note_10}
u(c_{1t},c_{2t+1})=\ln c_{1t}+\ln c_{2t+1}
\end{equation}
where we assume  no time discount for simplicity. In Appendix \ref{sec_robustness} we introduce a time discount and show that the main results hold.
Investors choose a saving rate to maximize life-time utility
\[
\max_{s^\ell_t \in [0,1]}\ln(1-s^\ell_t)w_t+\ln s^\ell_tw_tr_{t+1},
\]
which can be rewritten as 
$\ln U^{\ell}+\ln (w_t^2 r_{t+1})$ where
\begin{equation}\label{eq_10_optimal}
U^{\ell}:=\max_{s_t \in [0,1]}\{(1-s^\ell_t)s^\ell_t\}.
\end{equation}
The optimal saving rate of investors is $s_t^{\ell}=1/2$ and $U^{\ell}=1/4$. 
Entrepreneurs choose a saving rate to maximize  life-time utility
\[
\max_{s^b_t \in [0,1]}\left\{\ln(1-s^b_t)w_t+\ln (R f^{\prime}(k_{t+1})-(1-s^b_tw_t)r_{t+1})\middle|\ s^b_t \geq \frac{1-\lambda \phi_{t+1}}{w_t}\right\},
\]%
which can be rewritten as 
$\ln U^{b}(w_t,\phi_{t+1})+\ln (w_t^2 r_{t+1})$ where
\begin{equation}\label{eq_30_optimal}
U^{b}(w_t,\phi_{t+1},\lambda):=\max_{s^b_t \in [0,1]}\left\{\left(1-s^b_t\right)\left(\frac{\phi_{t+1}-1}{w_t}+s^b_t\right)\middle|\ s^b_t \geq \frac{1-\lambda \phi_{t+1}}{w_t}\right\}.
\end{equation}
The next proposition establishes the optimal saving of entrepreneurs.
\begin{proposition}\label{prop_optimal1}
For a given $\left(w_t,\phi_{t+1},\lambda\right)$, the optimal saving rate of entrepreneurs is
\begin{equation}\label{eq_130_optimal}
  s^{b}_t=\max\left\{\frac{1}{2}\left(1-\frac{\phi_{t+1}-1}{w_t}\right),\frac{1-\lambda \phi_{t+1}}{w_t}\right\}
\end{equation}
and
\begin{equation}\label{eq_230_optimal}
U^{b}(w_t,\phi_{t+1},\lambda)=
\begin{cases}
\frac{1}{4}\left(1+\frac{\phi_{t+1}-1}{w_t}\right)^2 & \textrm{if $w_t \geq 1-(2\lambda-1)\phi_{t+1}$}  \\[1.ex]
\left(1-\frac{1-\lambda \phi_{t+1}}{w_t}\right)\frac{(1-\lambda) \phi_{t+1}}{w_t} & \textrm{if  $1-\lambda\phi_{t+1}\leq w_t <1-(2\lambda-1)\phi_{t+1}$}.
\end{cases}
\end{equation}
If $w_t<1-\lambda\phi_{t+1}$, then young agents cannot become entrepreneurs because the credit constraint is violated even if they save their entire wage.
\end{proposition}

Proof of Proposition \ref{prop_optimal1} can be found in Appendix \ref{sec_appendix}.

\subsection{The Entrepreneurial Rent}

When making the saving decision, young agents take the value of $(w_t,\phi_{t+1},\lambda)$ as given and compare the value of $U^{\ell}$ and $U^{b}(w_t,\phi_{t+1},\lambda)$. If $U^{b}(w_t,\phi_{t+1},\lambda)<U^{\ell}$, then all young agents strictly prefer to become investors. If $U^{b}(w_t,\phi_{t+1},\lambda)>U^{\ell}$ then all young agents strictly prefer to become entrepreneurs. Therefore, it must be that $U^{b}(w_t,\phi_{t+1},\lambda)=U^{\ell}$ in equilibrium so that young agents are indifferent between becoming an entrepreneur and an investor. In equilibrium, some young agents become investors by lending $s^{\ell}w_t$ and the others become entrepreneurs by borrowing $1-s_t^b w_t$  in the competitive credit market. The next proposition establishes the entrepreneurial rent, which must hold in equilibrium. 
\begin{proposition}\label{prop_optimal2}
For a given $(w_t,\lambda)$, young agents are indifferent between becoming an entrepreneur and an investor if and only if  
\begin{equation}\label{eq_330_optimal}
\phi_{t+1}=\phi\left(w_t,\lambda\right):=
\begin{cases}
\frac{1}{2\lambda}\left(1-w_t+\sqrt{1-2 w_t+\frac{w_t^2}{1- \lambda}}\right) & \textrm{if  $w_t<2(1-\lambda)$} \\
1            & \textrm{if $w_t\geq 2(1-\lambda)$},
\end{cases}
\end{equation}
which is the solution to $U^{b}(w_t,\phi_{t+1},\lambda)=U^{\ell}$.
\end{proposition}

Proof of Proposition \ref{prop_optimal2} can be found in Appendix \ref{sec_appendix}.   From (\ref{eq_130_optimal}) and (\ref{eq_330_optimal}) we can see that if $\phi_{t+1}=1$, then the saving behavior of investors and entrepreneurs is identical and they consume the same amount of the final commodity when old. However, if $\phi_{t+1}>1$ then entrepreneurs save more than investors (entrepreneurs consume more than investors when old).
In either case, young agents will have to adjust the entrepreneurial rent toward the equilibrium level in order to maximize the lifetime utility and at the same time eliminate the excess demand/supply in the credit market. In other words, the equilibrium rent is the highest rent that entrepreneurs can earn and still be able to borrow all that they need, with no surplus or shortage in the credit market.

\begin{figure}[ht!]
	\centering
	\subfigure[$\phi_{t+1}=\phi(w_t,\lambda)$]
	{
		\label{fig:phi}
		\includegraphics[width=0.45\textwidth]{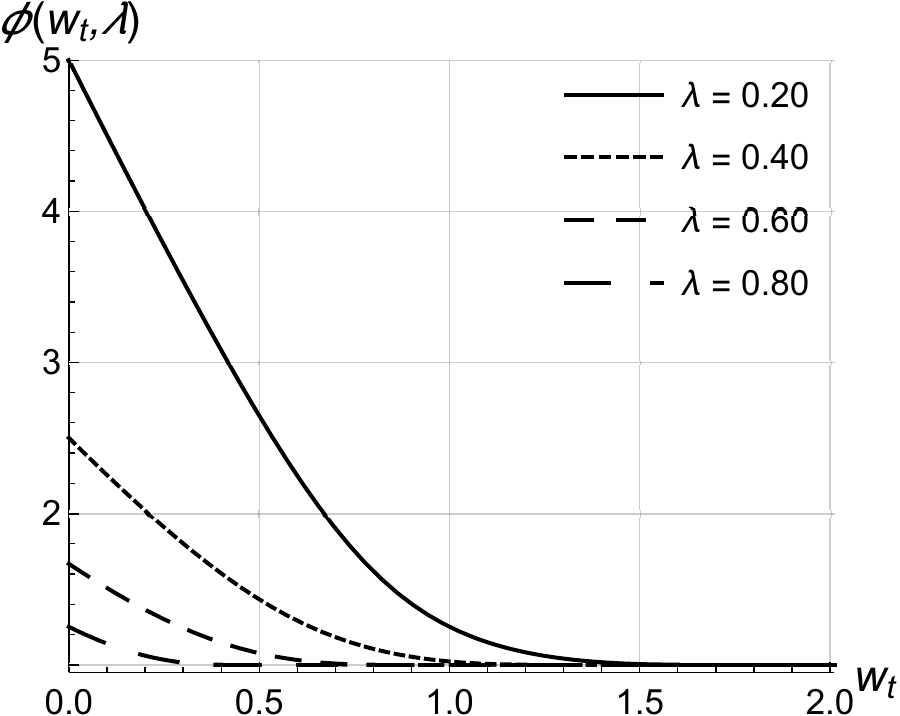}
	}
	\hspace{0.20cm}
	\subfigure[$s_t^b=s^b(w_t,\lambda)$]
	{
		\label{fig:sb}
		\includegraphics[width=0.45\textwidth]{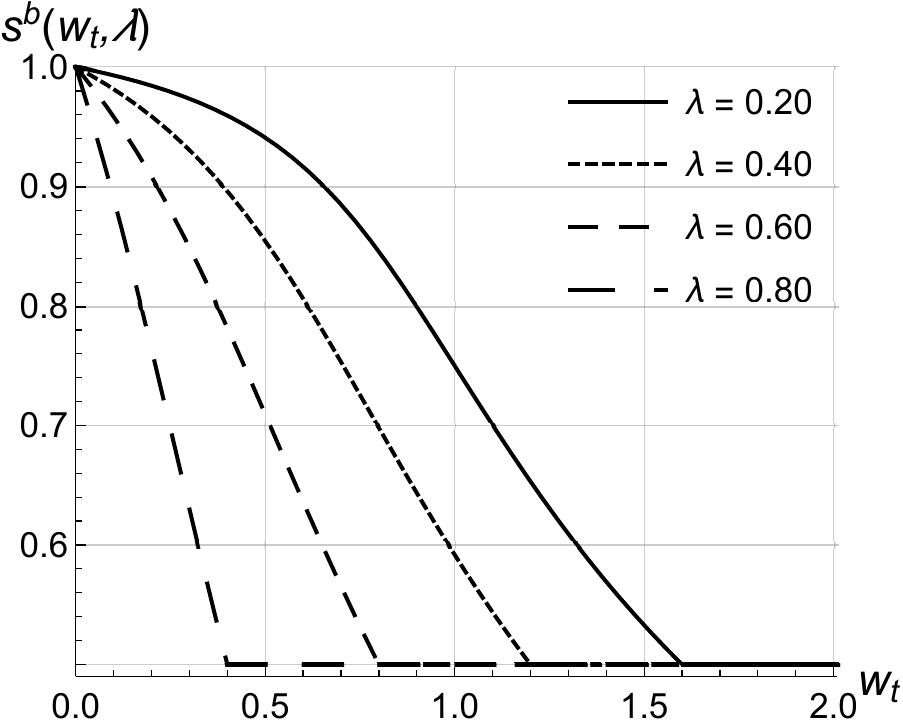}
	}
	
	\caption{The entrepreneurial rent ($\phi_{t+1}$) and the saving rate of entrepreneurs ($s_t^b$) for different values of parameter $\lambda$.}\label{fig:Figure_100}
	
\end{figure}

The entrepreneurial rent declines if the wage increases or the credit constraint is looser ($\lambda$ increases) as shown in Lemma \ref{lemma_theta}. If $0<w_t<2(1-\lambda)$, then $\phi_{t+1}=\phi(w_t,\lambda)>1$ and entrepreneurs and investors achieve the same level of lifetime utility despite the fact that they have different saving rates. Figure \ref{fig:phi} shows the entrepreneurial rent for different values of $\lambda$.

Substituting (\ref{eq_330_optimal}) into (\ref{eq_130_optimal}) we obtain the saving rate of entrepreneurs in equilibrium 
\begin{equation}\label{eq_20_sec_equilibrium}
s_t^b=s^{b}(w_t,\lambda):=
\begin{cases}
\frac{1-\lambda \phi(w_t,\lambda)}{w_t} & \textrm{if $ w_t<2(1-\lambda)$}      \\[1.ex]
\frac{1}{2}                               & \textrm{if $ w_t \geq 2(1-\lambda)$}.
\end{cases}
\end{equation}
The saving rate of entrepreneurs $s^b(w_t,\lambda)$ declines as the wage increases or the credit constraint is looser (see Lemma \ref{lemma_sb}). 
Figure \ref{fig:sb} shows the saving rate for different values of $\lambda$. 
If $0<w_t<2(1-\lambda)$, then from (\ref{eq_330_optimal}) and (\ref{eq_20_sec_equilibrium}) $\phi_{t+1}>1$ and $s_t^b>s_t^{\ell}=1/2$. Moreover, when the credit constraint is tighter, the entrepreneurial rent is higher and the saving rate of entrepreneurs is higher; entrepreneurs rely less on external funds. This captures the idea that ``anybody can make it through thrift.'' The young, who want to become entrepreneurs and face the credit constraint, save more in order to earn entrepreneurial rents in the future.\footnote{The main reasons, identified in the literature, why potential entrepreneurs save more than other investors are: (1) to accumulate the minimal capital requirements needed to engage in entrepreneurship and to implement projects as in our paper; (2) to hedge against uninsurable entrepreneurial risks; or (3) to cover the cost of external financing as in \cite{ghatak2001occupational}.}
We can show that $s_t^bw_t$ increases as the wage increases. This implies that by restricting to $w_t \in (0,2)$, we can guarantee $0<s_t^b w_t<1$ as $s^b_t=1/2$ when $w_t=2$. Hence, there is always a positive demand for credit; entrepreneurs always borrow in the credit market as it has been assumed (see Lemma \ref{lemma_sb}). 
\subsection{The National Saving Rate }

In equilibrium there will be only as many entrepreneurs as there are resources available divided by the final commodity each entrepreneur requires for investment projects. 
Let $s_t$ denote the national saving rate at time $t$. Since the resources available for the investment projects are the national saving $s_tw_t$ and each entrepreneur requires one unit of the final commodity for investment projects, the equilibrium size (i.e., the mass) of young agents who become entrepreneurs is $s_tw_t$ while the equilibrium size of young agents who become lenders is $1-s_tw_t$. 
Since each entrepreneur borrows $1-s_t^{b}w_t$ and each investor lends $s^{\ell}w_t=\frac{w_t}{2}$ units of the final commodity, the credit market clearing condition is 
\begin{equation}\label{eq_10_sec_equilibrium}
 s_t w_t\left(1-s_t^{b}w_t\right)=\left(1-s_t w_t\right)\frac{w_t}{2}.
\end{equation}
Substituting (\ref{eq_20_sec_equilibrium}) into (\ref{eq_10_sec_equilibrium}) we obtain the national saving rate in equilibrium   
\begin{equation}\label{eq_30_sec_equilibrium}
s_t=s(w_t,\lambda):=
\begin{cases}
\frac{1}{w_t+2\lambda \phi(w_t,\lambda)}  & \textrm{if  $w_t<2(1-\lambda)$}      \\[1.00ex]
\frac{1}{2}                               & \textrm{if  $w_t \geq 2(1-\lambda)$}.
\end{cases}
\end{equation}
The national saving rate increases on $w_t \in (0,1-\lambda)$, decreases on $w_t \in (1-\lambda,2(1-\lambda))$, and is constant at $1/2$ for $w_t \in (2(1-\lambda),2)$ leading to a hump-shaped saving rate with respect to the wage/wealth (see Lemma \ref{lemma_s_aggregate}). To understand this relationship we note that the wage affects the saving rate of entrepreneurs as well as the number of entrepreneurs (the saving rate of investors is constant).  Let $\pi_t:=s_tw_t\in(0,1)$ denote the fraction of entrepreneurs. We can rewrite (\ref{eq_10_sec_equilibrium}) as 
\begin{equation}\label{eq_40_sec_equilibrium}
 s_t=\pi_{t}\left(s_{t}^b-\frac{1}{2}\right)+\frac{1}{2}.
\end{equation}

\begin{figure}%[ht!]
	\centering
	\subfigure[$s(w_t,\lambda)$ ]
	{
		\label{fig:s}
		\includegraphics[width=0.45\textwidth]{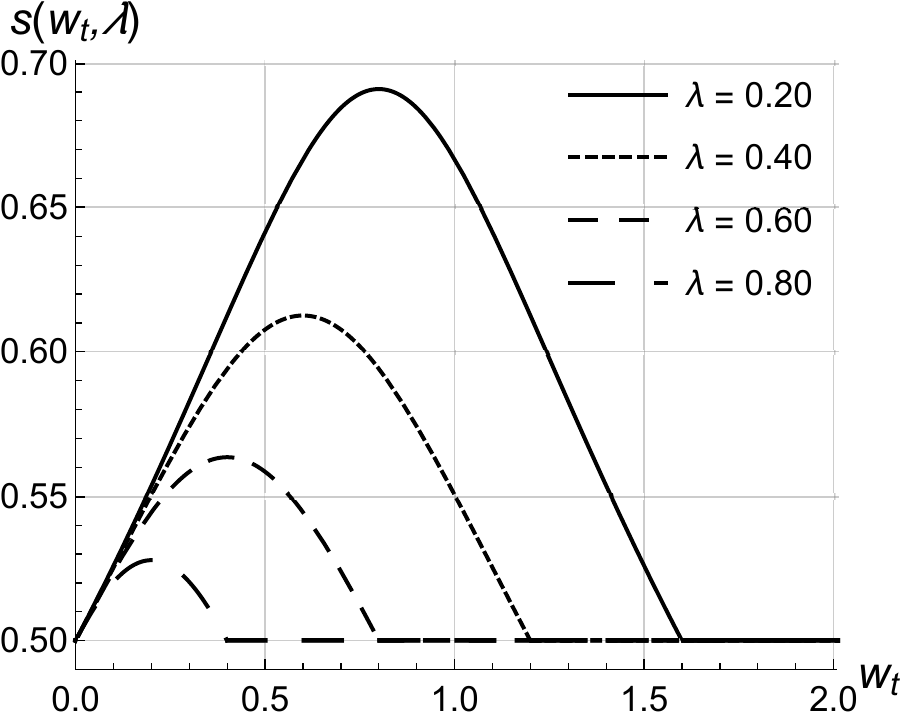}
	}
	\hspace{0.10cm}
	\subfigure[$\pi(w_t,\lambda)$ ]
	{
		\label{fig:pi}
		\includegraphics[width=0.45\textwidth]{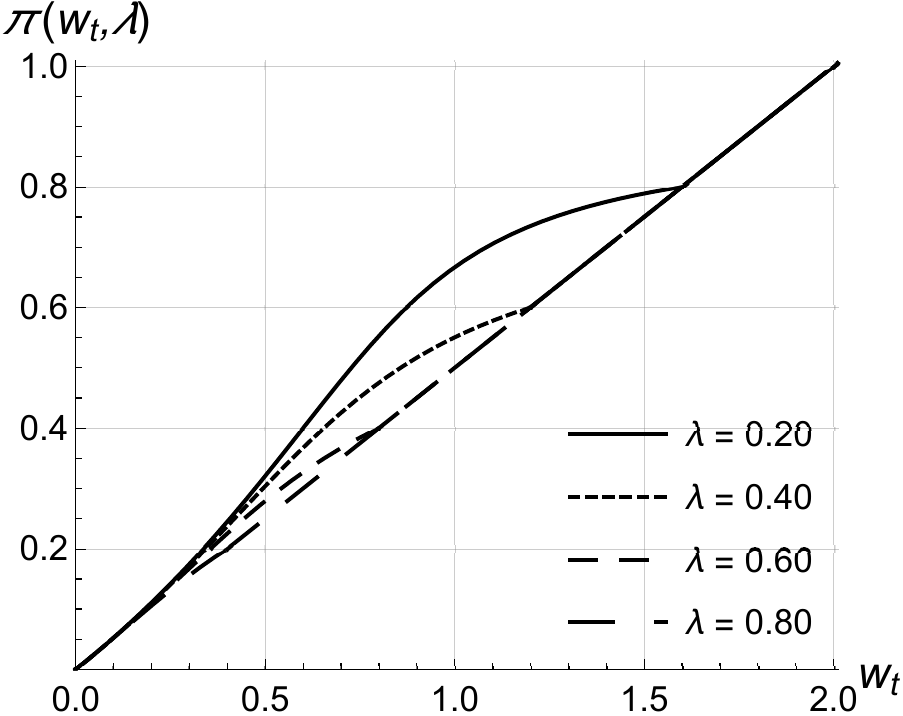}
	}
	
	\caption{The national saving rate $s_t$ and the fraction of entrepreneurs $\pi_t$ for different values of $\lambda$.}\label{fig:s-pi}
\end{figure}

The fraction of entrepreneurs $\pi_{t}$ increases but the saving rate of entrepreneurs $s_{t}^b$ decreases as the wage increases.\footnote{Lemma \ref{lemma_s_aggregate} demonstrates that  $\pi(w,\lambda)$ is strictly increasing in $w$ where $\pi(w,\lambda)=s(w,\lambda)w$.} Since the saving rate of entrepreneurs is higher than that of investors, the first effect causes the saving rate $s_t$ to rise. However, the second effect causes it to fall. The first effect dominates when the wage is low but the second effect eventually dominates. Once the wage is high enough so that the credit constraint no longer binds, $s_t^b=1/2$ and $s_t=1/2$. Figure \ref{fig:s} and Figure \ref{fig:pi} show the saving rate and the fraction of entrepreneurs respectively for different values of parameter $\lambda$. We can see that $\pi(w_t,\lambda)=s(w_t,\lambda)w_t$ is 
 strictly increasing in $w_t$ and strictly decreasing in $\lambda$ 
(see Lemma \ref{lemma_s_aggregate}). The fraction of entrepreneurs increases when the wage increases as they need to rely less on external funds. This increases the national savings. 

\subsection{The Transitional Wage Dynamics}

We are now ready to describe the transitional dynamics of the wage. With full depreciation of capital within a period, $k_{t+1}=R\pi(w_t,\lambda)$ and the evolution of the wage is given by
\begin{equation}\label{eq_system}
w_{t+1}=w(R\pi(w_t,\lambda)).
\end{equation}
Two important properties of the wage function should be highlighted here. First, strict concavity of $f$ implies strict monotonicity of $w$ because $w^{\prime}(k)=-kf^{\prime\prime}(k)>0$. Second, $w(0)=0$ because $f(0)=0$. These properties on the one hand imply the existence of $R^{+}>0$ such that $w(R)<2$ for $R \in (0,R^{+})$ and on the other hand imply that if $R \in (0,R^{+})$ and $w_{t} \in (0,2)$ then $\pi(w_{t},\lambda) \in (0,1)$ and $w_{t+1}=w[R\pi(w_{t},\lambda)] \in (0,w(R)) \subset(0,2)$. Hence, (\ref{eq_system}) defines a dynamical system on the state space $[0,2]$. Under (\ref{eq_system}) entrepreneurs always need to borrow in order to start an investment project. The following definition of equilibrium reconciles market clearing and individual optimality.
\begin{definition}\label{defn_equilibrium}
For a given parameter pair $(\lambda,R) \in (0,1) \times (0,R^+)$ and for a given initial wage $w_0=w(k_{0}) \in (0,2)$, the transitional dynamics is a equilibrium sequence $\{w_{t},s^b_{t},s_{t}\}_{t=0}^{\infty}$ such that in every period
\begin{enumerate}
\item The entrepreneurial rent adjusts so that young agents are indifferent between becoming an entrepreneur and becoming an investor; (\ref{eq_20_sec_equilibrium}) holds.
\item The fraction of entrepreneurs adjusts so that the credit market clears; (\ref{eq_10_sec_equilibrium}) holds. 
\item The fraction of entrepreneurs determines the wage in the next period; (\ref{eq_system}) holds.
\end{enumerate}
\end{definition}

We can easily verify that $w=0$ is the corner steady state because $\pi(0,\lambda)=0$ and $w(0)=0$. In general multiple interior steady states may exist. When multiple steady states exist, the initial condition determines to which steady state the economy converges, however the equilibrium sequence, $\{w_t\}^\infty_{t=0}$, of wages always forms a monotonic sequence for any $w_0 \in (0,2)$. The capital stock and the output per capita also form a monotonic sequence because $f$ and $w$ are both strictly increasing functions and $k_t=w^{-1}(w_t)$ and $y_t=f[w^{-1}(w_t)]$.\footnote{For example, if the production function is Cobb-Douglas, $f(k_t)=Ak_t^{\alpha}$ where $A>0$ is the total factor productivity and $\alpha \in (0,1)$ is the capital share, $w_t=(1-\alpha)y_t$.}  Given the monotone relationship between $w_t$ and $y_t$, we obtain 
\begin{theorem}The national saving rate $s(w_t, \lambda)$
	\begin{enumerate}
		\item  first increases and then decreases with output per capita $y_t$.
		
		\item  is higher when the credit constraint is tighter ($\lambda$ takes a lower value).
	\end{enumerate}
	\label{theorem:savingrate}
\end{theorem}
The theorem is consistent with the main empirical results that GDP per capita 
growth increases the national saving rate in poor countries while the opposite is true in rich countries, and that 
the growth of the credit-to-GDP ratio decreases the national saving rate.   
It is worthwhile to highlight here that the predictions in Theorem 1 hold even in an open economy, in which international credit and lending are allowed. 
In fact, the saving behavior of entrepreneurs depends on the wage and the severity of the credit constraint only and is independent of the interest rate. Therefore, we only need to confirm that the wage forms a monotonic sequence in a small open economy. In a small open economy the capital accumulation is determined by 
\[
k_{t+1}=\Phi(w_t)=(f')^{-1}\left(\frac{r^*\phi(w_t,\lambda)}{R}\right),
\]
where $r^*$ is the world interest rate. Note that $\Phi$ is a non-decreasing function in $w_t$ since $(f')^{-1}$ is a strictly decreasing function while $\phi$ is nonincreasing in $w_t$. Hence,  $w_{t+1}=W[\Phi(w_t)]$ implies that the equilibrium
wage sequence, $\{w_t\}_{t=0}^{\infty}$, as in the closed economy, always forms a monotonic
sequence for any $w_0\in(0,2)$. 

The theorem predicts the equilibrium relationships between the national saving rate, output, and the credit constraint. The credit constraint has a positive effect on the national saving rate because it encourages entrepreneurial saving similarly as in \cite{jappelli1994saving} where it reduces consumption. However, previous models such as \cite{jappelli1994saving} do not generate our main result that the national saving rate first increases and then decreases as output per capita increases.

\subsection{Motivation of the Econometric Model Specification\label{sec:specification}}
 
We show how the econometric specification in (\ref{eq:12}) and (\ref{eq:interactions}) can be derived from our model.\footnote{There is a fraction of the population that works (called the young in the OLG model) and a fraction of the population that is retired (called the old in the OLG model). For estimation of the econometric model, we need variation in the data for the variables of interest; i.e. variation that comes from the cross-section and variation that comes from the time-series. Regarding the time-series dimension, the panel data we use are at an annual frequency; this maximizes observations. 
In any given year, there is a fraction of the population that works and a fraction of the population that is retired.  
}
Taking logarithms we obtain $\ln s_t=\ln s((w\circ f^{-1})(y_{t}),\lambda_{t})$. For small values of ${\triangle}{\ln}y_{t}={\ln}y_{t}-{\ln}y_{t-1}$ and ${\triangle}{\ln}\lambda_{t}={\ln}\lambda_{t} -{\ln}\lambda_{t-1}$, ${\triangle}{\ln}s_{t}={\ln}s_{t}-{\ln}s_{t-1}$ can be approximated as 
\begin{equation}
{\triangle}{\ln}s_{t}\approx\gamma_{t-1}{\triangle}{\ln}y_{t}+\theta_{t-1}{\triangle}{\ln}\lambda_{t}\label{eq80}
\end{equation}
 where $\gamma_{t-1}:=\frac{w_{t-1}s_1(w_{t-1},\lambda_{t-1})}{s(w_{t-1},\lambda_{t-1})}\frac{y_{t-1}(w\circ f^{-1})^{\prime}(y_{t-1})}{(w\circ f^{-1})(y_{t-1})}$ and $\theta_{t-1}:=\frac{w_{t-1}s_2(w_{t-1},\lambda_{t-1})}{s(w_{t-1},\lambda_{t-1})}$.\footnote{Observe $\ln{g(x_{t})} \approx \ln{g(x_{t-1})}+\frac{g^{\prime}(x_{t-1})}{g(x_{t-1})}(x_{t}-x_{t-1})$. Since $\Delta\ln{x_{t}}=\ln{x_{t}}-\ln{x_{t-1}}\approx\frac{x_{t}-x_{t-1}}{x_{t-1}}$, it follows that $\ln{g(x_{t})}\approx\ln{g(x_{t-1})}+\frac{x_{t-1}g^{\prime}(x_{t-1})}{g(x_{t-1})}\Delta\ln{x_{t}}$.}
 Proof of Lemma \ref{lemma_s_aggregate} in Appendix \ref{sec_appendix} derives both elasticities. The sign of the income elasticity of the national saving rate changes from positive to negative  at $w=1-\lambda$. The elasticity of the saving rate with respect to the parameter $\lambda$ is always negative. If the production function is Cobb-Douglas, then $\frac{y(w\circ f^{-1})^{\prime}(y)}{(w\circ f^{-1})(y)}=1$.
 
 Suppose that the parameters do not vary significantly over time (we test this hypothesis in Section \ref{sec:empirical-results}). Then, we can write (\ref{eq80}) for a panel of $i=1...N$ countries as
 \begin{equation}
 \triangle\ln s_{it}\approx\gamma_{i}\triangle\ln y_{it}+\theta_{i}\triangle\ln\lambda_{it}.\label{eq80-1}
 \end{equation}
 The above equation shows that a within-country change in the log of the national saving rate should be related to a within-country change in the log of $y$ and $\lambda.$ In (\ref{eq:12}), $\gamma$ and $\theta$ thus capture the average elasticity effect of a marginal within-country change in GDP per capita and in the credit-to-GDP ratio on a change in the national saving rate. In other words, $\gamma$ and $\theta$ are sample mean elasticity effects. We check in Section \ref{sec:empirical-results} whether our restricted panel data model in (\ref{eq:12}) consistently estimates these average elasticity effects, using the \cite{pesaran1995estimating} MG estimator.
 
We note that for a relatively low value of $y_{i}$, $w_i=(w\circ f^{-1})(y_{i})<1-\lambda$ while $w_i=(w\circ f^{-1})(y_{i})>1-\lambda$ for a relatively high value of $y_{i}$. It follows that $\gamma_{i}$ is positive (negative) for a low (high) value of $y_{i}$ and any fixed value of $\lambda_{i}$ because $\gamma_i=\frac{w_is_1(w_i,\lambda_i)}{s(w_i,\lambda_i)}\frac{y_i(w\circ f^{-1})^{\prime}(y_i)}{(w\circ f^{-1})(y_i)}>0$ when $w_i \in (0,1-\lambda_i)$ and $\gamma_i=\frac{w_is_1(w_i,\lambda_i)}{s(w_i,\lambda_i)}\frac{y_i(w\circ f^{-1})^{\prime}(y_i)}{(w\circ f^{-1})(y_i)}<0$ when $w_i \in (1-\lambda_i,2(1-\lambda_i))$. The income elasticity of the national saving rate can therefore be either positive or negative. On the other hand, since $\theta_{i}=\frac{w_is_2(w_i,\lambda_i)}{s(w_i,\lambda_i)}<0,$ it follows that $\theta_{i}$ is negative for any fixed values of $\lambda_{i}$, $y_{i}$, and $w_i=(w\circ f^{-1})(y_{i})$. Hence, our model unambiguously predicts a negative elasticity effect of a higher credit-to-GDP ratio on the national saving rate.

Let $F(y,\lambda)\equiv \frac{(w\circ{f}^{-1})(y)s_1((w\circ{f}^{-1})(y),\lambda)}{s((w\circ{f}^{-1})(y),\lambda)}\frac{y(w\circ{f}^{-1})^{\prime}(y)}{(w\circ{f}^{-1})(y)}$. Then the functional form of the interaction model can be derived from our model by applying the following first-order Taylor expansion
\[
\gamma_{i}\equiv F(y_{i},\lambda_{i})\approx F(\hat{y},\hat{\lambda})+F_{1}(\hat{y},\hat{\lambda})(y_{i}-\hat{y})+F_{2}(\hat{y},\hat{\lambda})(\lambda_{i}-\hat{\lambda}),
\]
where $\hat{y}$ is the unique solution to $(w\circ{f}^{-1})(y)=1-\lambda$ and $\hat{\lambda}$ is the cross-country average of $\lambda_{i}$. Substituting the above expression in (\ref{eq80-1}) yields the following relationship between the change in the log of the national saving rate and the change in the log of GDP per capita  and the change in the log of the credit-to-GDP ratio
\[
\triangle\ln s_{it}=\left(\gamma^{\prime}+\delta\lambda_{i}+\zeta y_{i}\right)\triangle\ln y_{it}+\theta'\triangle\ln\lambda_{it}.
\]
where $F(\hat{y},\hat{\lambda})=0$, $\gamma^{\prime}\equiv-\hat{y}F_{1}(\hat{y},\hat{\lambda})-\hat{\lambda}F_{2}(\hat{y},\hat{\lambda})$, $\zeta\equiv F_{1}(\hat{y},\hat{\lambda})$ and $\delta\equiv F_{2}(\hat{y},\hat{\lambda})$. Note that $F_{1}(\hat{y},\hat{\lambda})<0$ and $F_{2}(\hat{y},\hat{\lambda})<0$. This implies that $\gamma^{\prime}>0$, $\delta<0$, $\zeta<0$ and $\theta^{\prime}<0$ as before. Hence, our model predicts that the income elasticity of the national saving rate is larger in countries with lower GDP per capita and a lower credit-to-GDP ratio.

\section{Empirical Support for the Mechanism\label{sec:support-mechnism}}

The theoretical model in Section \ref{sec_model} explains the empirical findings that the national saving rate follows a hump-shaped transitional dynamics by the interplay of entrepreneurial savings at the intensive and extensive margins. The model predicts that when the wealth increases, more agents can become entrepreneurs but entrepreneurs save less, and when the wealth is low, the extensive margin dominates the intensive margin; the opposite is true when the wealth is high. This section examines the mechanism of the theoretical model by utilizing cross-country time series data of 
the number of new businesses registered as a proxy for the extensive margin and the corporate saving rate as a proxy for the intensive margin.

\subsection{New Businesses Registered}

In this section we estimate the relationship between the
growth rate of the number of new businesses registered and GDP per
capita growth (constant price PPP based). We also estimate
the relationship between the growth rate of the number of new businesses
registered and the growth rate of the credit-to-GDP ratio. We obtained
data on the number of new businesses registered from the World Development
Indicators \cite{bank2020world}. Specifically, the variable we use
is the number of new limited liability corporations registered in
the calendar year (per 1000 people aged 15 to 64). These data are
available annually from 2006 onwards.

\begin{table}[ht!]
	\singlespacing
	\footnotesize
	
	\begin{center}
		\begin{tabular}{llll}
			\hline
			
			\multicolumn{4}{c}{$\Delta\ln(\text{New Businesses Registered}_{it})$} \\
			
			\hline
			& (1)        & (2)   & (3)  \\

			$\Delta\ln(y_{it}) $    & 1.994***    & 2.417*** & 2.605*** \\
			
			& (0.319)&(0.391)    & (0.394)\\

			$\Delta\ln(y_{it})*y_i$  & & -0.024*   & -0.035** \\
			
			& & (0.013)& (0.014)\\
			
			$\Delta\ln(\lambda_{it})$ &    &  & 0.226** \\
			
			& & &(0.104)\\
			
			$\Delta\ln(\lambda_{it})*y_i$      &  &    & -0.012** \\
			
			& &  &(0.005)\\

			Country FE         &  Yes       & Yes           & Yes       \\
			
			Year FE             & Yes        & Yes          & Yes       \\
			
			Observations       & 810       & 810          & 810   \\
			
			\hline
		\end{tabular}
	\end{center}
	
	Note: The method of estimation is least squares. Standard errors are shown in parentheses.. The dependent variable is the change in the log of the new business registered per 1000 people aged between 15 to 64. *Significantly different from zero at the 10 percent significance level, **5 percent significance level, ***1 percent significance level.\vspace{1em}
	\caption{\label{tab:business}The relationship between GDP per capita growth, the growth rate of the credit-to-GDP ratio and the growth rate of new businesses registered
	}
\end{table}

Column (1) of Table \ref{tab:business} shows least squares estimates
of the average relationship between the growth rate of the number
of new businesses registered and GDP per capita growth. The panel
covers 105 countries over the period 2006-2017. The dependent variable
is the year $t-1$ to $t$ change in the log of new businesses registered
per capita and the right-hand-side variable is the year $t-1$ to
$t$ change of the log of GDP per capita. Control variables are country
and year fixed effects.

From column (1) of Table \ref{tab:business}, one can see that on
average there is a significant positive relationship between GDP per
capita growth and the growth rate of new businesses registered. The
estimated coefficient on the $t-1$ to $t$ change in the log of GDP
per capita, $\Delta\ln(y_{it})$, is around 1.9 and has a standard
error of around 0.3. One can reject the hypothesis that the estimated
coefficient is equal to zero at the 1 percent significance level.
Quantitatively, the estimate in column (1) of Table \ref{tab:business}
can be interpreted as follows: a 1 percentage point increase in GDP
per capita growth raises the growth rate of new businesses registered
by around 2 percentage points.

Column (2) of Table \ref{tab:business} shows that the effect of GDP
per capita growth on the growth rate of new business registered is
larger for poorer countries. This can be seen from the significant
negative coefficient on $\Delta\ln(y_{it})*y_{i}$, which is the interaction
between the $t-1$ to $t$ change in the log of GDP per capita and
countries' average GDP per capita (in the table, reported in 1000s
of dollars) over the period 2006-2017. Quantitatively, the estimates
in column (2) can be interpreted as follows. Consider a low income
country with average GDP per capita over the period 2006-2017 equal
to USD 1,000. The estimates in column (2) suggest that, for this country,
a 1 percentage point increase in GDP per capita growth raises the
growth rate of new businesses registered by around 2.4 percentage
points. Consider now a middle income country with average GDP per
capita from 2006 to 2017 equal to USD 10,000. The estimates in column
(2) suggest that, for this country, a 1 percentage point increase
in GDP per capita growth raises the growth rate of new businesses
registered by around 2.2 percentage points. For a high income country
with GDP per capita equal to USD 50,000 the effect is much smaller.
For that country, the estimates in column (2) of Table \ref{tab:business}
suggest that a 1 percentage point increase in GDP per capita growth
raises the growth rate of new businesses registered by around 1.4
percentage points.

Column (3) of Table \ref{tab:business} shows that the relationship
between the growth rate of new businesses registered and GDP per capita
growth is robust to controlling for the growth rate of the credit-to-GDP
ratio. From column (3) of Table \ref{tab:business} one can see that
the growth rate of the credit-to-GDP ratio has a significant positive
effect on the growth rate in new businesses registered, but less so,
the higher is the GDP per capita in the economy.

\subsection{The Corporate Saving Rate}

In this section we estimate the relationship between the
growth rate of the corporate saving rate and GDP per capita growth.
We also estimate the relationship between the growth rate
of the corporate saving rate and the growth rate of the credit-to-GDP
ratio. Our data on the gross saving rate of the corporate sector
are from \cite{chen2017global}. We compute the gross corporate saving
rate as the gross saving of the corporate sector divided by the gross
value added of the corporate sector. The data on GDP per capita (PPP
based) and the credit-to-GDP ratio (the GDP share of domestic credit
to the private sector) are from the World Development Indicators \cite{bank2020world}.

\begin{table}[ht!]
	\singlespacing
	\footnotesize
	
	\begin{center}
		\begin{tabular}{llll}
			\hline
			
			\multicolumn{4}{c}{$\Delta\ln(\text{Gross Corporate saving rate}_{it})$} \\
			
			\hline
			& (1)        & (2)   & (3)  \\

			$\Delta\ln(y_{it}) $    & -0.621***    & -0.021 & -0.029 \\
			
			& (0.166)&(0.195)    & (0.079)\\

			$\Delta\ln(y_{it})*y_i$  & & -0.030***   & -0.028*** \\
			
			& & (0.005)& (0.004)\\
			
			$\Delta\ln(y_{it})*\lambda_{i}$ &    &  & -0.025** \\
			
			& & &(0.010)\\
			
			$\Delta\ln(\lambda_{it})$      &  &    & -0.015 \\
			
			& &  &(0.009)\\

			Country FE         &  Yes       & Yes           & Yes       \\
			
			Year FE             & Yes        & Yes          & Yes       \\
			
			Observations       & 887       & 887          & 706   \\
			
			\hline
		\end{tabular}
	\end{center}
	
	Note: The method of estimation is least squares. Standard errors are shown in parentheses. The dependent variable is the change in the log of gross corporate saving rate. *Significantly different from zero at the 10 percent significance level, ** 5 percent significance level, *** 1 percent significance level.\vspace{1em}
	\caption{\label{tab:coporate_saving}The relationship between GDP per capita growth, the growth rate of credit-to-GDP ratio  and the growth rate of the corporate saving rate
	}
\end{table}

Column (1) of Table \ref{tab:coporate_saving} shows least squares
estimates of the average the relationship between the growth rate
of the corporate saving rate and GDP per capita growth. The panel
covers 59 countries over the period 1992-2013. The dependent variable
is the year $t-1$ to $t$ change in the log of the corporate saving
rate and the right-hand-side variable is the year $t-1$ to $t$ change
of the log of GDP per capita. Control variables are country and year
fixed effects.

From column (1) of Table \ref{tab:coporate_saving}, one can see that
on average there is a significant negative relationship between GDP
per capita growth and the growth rate of the corporate saving rate.
The estimated coefficient on the $t-1$ to $t$ change in the log
of GDP per capita is around -0.6 and has a standard error of around
0.2. One can reject the hypothesis that the estimated coefficient
is equal to zero at the 1 percent significance level. Quantitatively,
the estimate in column (1) of Table \ref{tab:coporate_saving} can
be interpreted as follows: a 1 percentage point increase in GDP per
capita growth decreases the growth rate of the corporate saving rate
by around 0.6 percentage points, on average.

Column (2) of Table \ref{tab:coporate_saving} shows that the negative
effect of GDP per capita growth on the growth rate of the corporate
saving rate is larger (in absolute value) for richer countries. This
can be seen from the significant negative coefficient on $\Delta\ln(y_{it})*y_{i}$,
which is the interaction between the $t-1$ to $t$ change in the
log of GDP per capita and countries' average GDP per capita (in the
table, reported in USD 1000s) over the period 1992-2013. Quantitatively,
the estimates in column (2) can be interpreted as follows. Consider
a low income country with average GDP per capita over the period 1992-2013
equal to USD 1,000. The estimates in column (2) suggest that, for
this country, a 1 percentage point increase in GDP per capita growth
decreases the growth rate of the corporate saving rate by around
0.05 percentage points. Consider now a middle income country with
average GDP per capita over the period 1992-2013 equal to USD 10,000.
The estimates in column (2) suggest that, for this country, a 1 percentage
point increase in GDP per capita growth decreases the growth rate
of the corporate saving rate by around 0.3 percentage points. For
a high income country with GDP per capita equal to USD 50,000 the
effect is much larger. For that country, the estimates in column (2)
of Table \ref{tab:coporate_saving} suggest that a 1 percentage point
increase in GDP per capita growth decreases the growth rate of the
corporate saving rate by around 1.5 percentage points.

Column (3) of Table \ref{tab:coporate_saving} shows that GDP per
capita growth has a larger negative effect on the growth rate of the
corporate saving rate, the larger is the credit-to-GDP ratio. This
can be seen from the significant negative coefficient on $\Delta\ln(y_{it})*\lambda_{i}$,
which is the interaction between the $t-1$ to $t$ change in the
log of GDP per capita and countries' average credit-to-GDP ratio over
the period 1992-2016. The estimated coefficient on this interaction
term is -0.025 and has a standard error of 0.010. This coefficient
can be interpreted as follows. For each 10 percentage points increase
in the credit-to-GDP ratio, the effect of a 1 percentage point increase
in the growth rate of GDP per capita reduces the growth rate of the
corporate saving rate by an additional 0.25 percentage points.

\section{Conclusion\label{sec_conclusion}}

Based on panel data covering 130 countries over the period 1960-2017 we find that the income elasticity of the national  saving rate is significantly decreasing in GDP per capita and the credit-to-GDP ratio. There exists a threshold of  GDP per capita above which the income elasticity of the national saving rate is negative. The elasticity is significantly positive in low income countries.  For the majority of high income countries it is significantly negative. 
In addition, the elasticity decreases when the credit-to-GDP ratio is higher. So much so, that in countries with a low credit-to-GDP ratio GDP per capital growth increases the saving rate while in countries with a high credit-to-GDP ratio the opposite is the case.

To explain the empirical findings we build a model in which entrepreneurs are credit constrained and investment projects are indivisible. The credit constraint creates rents for entrepreneurs. The indivisible investment size does not permit all agents to obtain credit to finance entrepreneurial activities. This creates dynamic incentives for entrepreneurs to save more and rely less on external funds. The resulting saving behavior of entrepreneurs generates the relationship between GDP per capita growth, the national saving rate and the credit constraint. We present supporting evidence for our theoretical findings by utilizing cross-country time series data of the number of new businesses registered and the corporate saving rate.

\appendix

\section{A Cobb-Douglas Example\label{sec:cobb-douglas}}

Suppose that the production function is Cobb-Douglas, i.e, $f(k)=k^{\alpha}$ where   $\alpha \in (0,1)$. It follows that $w(k)=(1-\alpha)k^{\alpha}$, $R^{+}=(\frac{2}{1-\alpha})^{\frac{1}{\alpha}}$ and $w^{\prime}(0)=\infty$. This implies that  the corner steady state is always locally unstable and there exists either a unqiue interior steady state or an odd number of interior steady states, which solve $\Pi(w,\lambda)=R$ where $\Pi(w,\lambda)=\frac{w^{-1}(w)}{s(w,\lambda)w}$. 
If
\begin{equation}\label{eq_100_ss}
\frac{w \Pi_1(w,\lambda)}{\Pi(w,\lambda)}=\frac{1-\alpha}{\alpha}-\frac{w s_1(w,\lambda)}{s(w,\lambda)}>0
\end{equation}
i.e., if the elasticity of output is small relative to the elasticity of saving ($\alpha<1/2$ is sufficient),   
$\Pi(w,\lambda)$ is monotonically increasing in $w$ and thus there exists a unique interior steady state. Let $w^*(R,\lambda)$ denote the unique steady state. Suppose that  $w^*(R,\lambda)>1-\lambda$. If $w_0<1-\lambda$, then the saving rate $s_t$ first increases and then decreases as $w_t$ (or $y_t$) converges to the steady state in the long run.

\section{Remaining Proofs \label{sec_appendix}}

We eliminate time subscripts for notational convenience. 

\textbf{Proof of Proposition \ref{prop_optimal1}:} Let
\begin{equation}\label{eq_10_proof_of_prop_optimal1}
\begin{array}{ccc}
  s^b_1=\frac{1}{2}\left(1-\frac{\phi-1}{w}\right) & \textrm{and} &  s^b_2=\frac{1-\lambda \phi}{w}.
\end{array}
\end{equation}
We can easily verify that $s=s^b_1$ solves the  unconstrained optimization problem
of entrepreneurs%
\begin{equation}\label{eq_20_proof_of_prop_optimal1}
  U^b=\max_{s \in [0,1]}\left\{\left(1-s\right)\left(\frac{\phi-1}{w}+s\right)\right\}.
\end{equation}
If $w \geq 1-(2\lambda-1)\phi$, then $s^b_1 \geq s^b_2$ and thus entrepreneurs can overcome the credit constraint if their saving rate is $s_1^b$. In such case, $U^b=\frac{1}{4}(1+\frac{\phi-1}{w})^2$.

If $w \in \left[1-\lambda\phi,1-(2\lambda-1)\phi\right)$, then $s_1^b<s^b_2 \leq 1$ and thus entrepreneurs can overcome the credit constraint if their saving rate is $s_2^b$. In such case, $U^b=(1-\frac{1-\lambda \pi}{w})\frac{(1-\lambda)\phi}{w}$.

If $w<1-\lambda\phi$, then $s_1^b<1<s^b_2$ and thus  entrepreneurs cannot overcome the credit constraint even if they save their entire wage. 
\qed

\textbf{Proof of Proposition \ref{prop_optimal2}:} If $w \geq 1-(2\lambda-1)\phi$, then it follows from (\ref{eq_230_optimal}) that $U_t^b=U^{\ell}$ $\Leftrightarrow$ $\phi=1$. Hence, $w \geq 1-(2\lambda-1)\phi$ $\Leftrightarrow$ $w \geq 2(1-\lambda)$. If $w \in [1-\lambda \phi,1-(2\lambda-1)\phi)$, then it follows from (\ref{eq_230_optimal}) that $U_t^b=U^{\ell}$ $\Leftrightarrow$ $\phi=\frac{1}{2\lambda}\left(1-w+\sqrt{1-2w+\frac{w^2}{1-\lambda}}\right)$. Hence, $w \in [1-\lambda \phi,1-(2\lambda-1)\phi)$ $\Leftrightarrow$ $w \in [0,2(1-\lambda))$.
\qed

\begin{lemma}\label{lemma_theta} (a) For $\lambda \in (0,1)$, the entrepreneurial rent $\phi(w,\lambda)$ is a continuous and strictly decreasing function on $w \in (0,2(1-\lambda))$ and satisfies the following boundary properties
\begin{equation}\label{eq_100_lemma_theta}
\begin{array}{ccc}
\lim_{w \downarrow 0}\phi(w,\lambda)=\frac{1}{\lambda} & \text{and} & \lim_{w \uparrow 2(1-\lambda)}\phi(w,\lambda)=1.
\end{array}
\end{equation}
(b) For $w \in (0,2(1-\lambda))$, $\phi(w,\lambda)$ is a strictly decreasing function while $\lambda\phi(w,\lambda)$ is a strictly increasing function on $\lambda \in (0,1)$.
\end{lemma}

\textbf{Proof of Lemma \ref{lemma_theta}:} If $\lambda \in (0,1)$ and $w<2(1-\lambda)$, then the entrepreneurial rent is
\begin{equation}\label{eq_400_lemma_theta}
\begin{array}{ccc}
  \phi(w,\lambda)=\frac{1-w+\psi(w,\lambda)}{2\lambda} & \textrm{where} & \psi(w,\lambda):=\sqrt{1-2w+\frac{w^2}{1-\lambda}}.
\end{array}
\end{equation}

(a) Differentiating both sides of (\ref{eq_400_lemma_theta}) with respect to $w$ and re-arranging terms, we obtain 
\begin{equation}\label{eq_500_lemma_theta}
  \phi_1(w,\lambda)=
  \frac{1}{\psi(w,\lambda)}\left(\frac{w}{2(1-\lambda)}-\phi(w,\lambda)\right)<0
\end{equation}
because when $w \in (0,2(1-\lambda))$, $\frac{w}{2(1-\lambda)}<1<\phi(w,\lambda)$ and $\psi(w,\lambda) \in (1,1/\lambda)$. This implies monotonicity of $w \mapsto \phi(w,\lambda)$. Taking limits of both sides of (\ref{eq_400_lemma_theta}), we obtain the boundary properties of $\phi$, which along with $\phi(w,\lambda) \equiv 1$ for $w \geq 2(1-\lambda)$ imply continuity of $\phi$.

(b) Differentiating both sides of (\ref{eq_400_lemma_theta}) with respect to $\lambda$ and re-arranging terms, we obtain 
\begin{equation}\label{eq_600_lemma_theta}
 \ds\frac{\lambda \phi_2(w,\lambda)}{\phi(w,\lambda)}=\frac{w^2}{4(1-\lambda)^2}\frac{1}{\psi(w,\lambda)\phi(w,\lambda)}-1 \in (-1,0)
\end{equation}
because when $w \in (0,2(1-\lambda))$, $\frac{w}{2(1-\lambda)}<1<\phi(w,\lambda)$ and $\psi(w,\lambda) \in (1,1/\lambda)$. The monotonicity properties of $\lambda \mapsto \phi(w,\lambda)$ and $\lambda \mapsto \lambda\phi(w,\lambda)$ are implied by (\ref{eq_600_lemma_theta}).
\qed

\begin{lemma}\label{lemma_sb} (a) For $\lambda \in (0,1)$, the saving rate of entrepreneurs $s^b(w,\lambda)$ is a strictly decreasing function on $w \in (0,2(1-\lambda))$ and satisfies the following boundary properties
    \begin{equation}\label{eq_100_lemma_sb}
    \begin{array}{ccc}
    \lim_{w \downarrow 0}s^b(w,\lambda)=1 & \textrm{and} & \lim_{w \uparrow 2(1-\lambda)}s^b(w,\lambda)=\frac{1}{2}.
    \end{array}
    \end{equation}
(b) For  $w \in (0,2(1-\lambda))$, $s^b(w,\lambda)$ is a strictly decreasing function on $\lambda \in (0,1)$.
\end{lemma}

\textbf{Proof of Lemma \ref{lemma_sb}:} (a) In equilibrium $U^{b}=U^{\ell}$ $\Leftrightarrow$
\begin{equation}\label{eq_100_lemma_sb}
\begin{array}{ccc}
  \left(1-\frac{1}{w}+\frac{\lambda \phi(w,\lambda)}{w}\right)\frac{(1-\lambda)\phi(w,\lambda)}{w}=\frac{1}{4} & \Leftrightarrow & \frac{1-\lambda \phi(w,\lambda)}{w}=1-\frac{w}{4(1-\lambda)\phi(w,\lambda)}.
\end{array}
\end{equation}
Monotonicity and boundary properties of $w \mapsto \phi(w,\lambda)$ with (\ref{eq_100_lemma_sb}) imply monotonicity and boundary properties of $w \mapsto s^b(w,\lambda)$.

(b) Monotonicity of $\lambda \mapsto s^b(w,\lambda)$ follows from Lemma \ref{lemma_theta}.
\qed

\begin{lemma}\label{lemma_s_aggregate} (a) For $\lambda \in (0,1)$, the national saving rate
\begin{equation}\label{eq_50_lemma_s_aggregate}
s(w,\lambda) \equiv \begin{cases}
 \frac{1}{w+2\lambda\phi(w,\lambda)}   & \textrm{if $w<2(1-\lambda)$}      \\[1.00ex]
 \frac{1}{2}                           & \textrm{if $w \geq 2(1-\lambda)$}
\end{cases}
\end{equation}
first increases and then decreases on $w \in(0,2(1-\lambda))$ achieving its local maximum at $w=1-\lambda$ and satisfying the boundary properties
\begin{equation}\label{eq_100_lemma_s_aggregate}
\begin{array}{ccc}
  \lim_{w \downarrow 0}s(w,\lambda)=\lim_{w \uparrow 2(1-\lambda)}s(w,\lambda)=\frac{1}{2} & \textrm{and} & \lim_{w \rightarrow 1-\lambda}s(w,\lambda)=\frac{1}{\lambda}.
\end{array}
\end{equation}

(b) For $\lambda \in (0,1)$, the fraction of entrepreneurs $\pi(w,\lambda)=s(w,\lambda)w$ is an increasing function on $w>0$ satisfying the boundary properties
\begin{equation}\label{eq_200_lemma_s_aggregate}
\begin{array}{ccc}
  \lim_{w \downarrow 0}\pi(w,\lambda)=\frac{1}{2} & \textrm{and} & \lim_{w \uparrow 2(1-\lambda)}\pi(w,\lambda)=\frac{\lambda}{2}.
\end{array}
\end{equation}

(c) For $w \in (0,2(1-\lambda))$, $s(w,\lambda)$ and $\pi(w,\lambda)$ are both strictly decreasing functions on $\lambda \in (0,1)$.
\end{lemma}

\textbf{Proof of Lemma \ref{lemma_s_aggregate}:} (a) It follows from (\ref{eq_400_lemma_theta}) and (\ref{eq_50_lemma_s_aggregate}) that
\begin{equation}\label{eq_300_lemma_s_aggregate}
s(w,\lambda)=\ds\frac{1}{w+2\lambda\phi(w,\lambda)}=\ds\frac{1}{1+\psi(w,\lambda)}
\end{equation}
where $\psi$ is defined in (\ref{eq_400_lemma_theta}). Differentiating both sides of (\ref{eq_300_lemma_s_aggregate}) and using the definition of $\psi$, we obtain 
\begin{equation}\label{eq_400_lemma_s_aggregate}
\begin{array}{ccc}
\frac{w s_1(w,\lambda)}{s(w,\lambda)}=\frac{[s(w,\lambda)]^2w}{1-s(w,\lambda)}\left(1-\frac{w}{1-\lambda}\right) & \textrm{and} & \frac{\lambda s_2(w,\lambda)}{s(w,\lambda)}=-\frac{\lambda [s(w,\lambda)]^2}{2(1-s(w,\lambda))}\left(\frac{w}{1-\lambda}\right)^2
\end{array}
\end{equation}
where $s_1(w,\lambda):=\frac{\partial s(w,\lambda)}{\partial w}$ and $s_2(w,\lambda):=\frac{\partial s(w,\lambda)}{\partial \lambda}$. This implies that $s(w,\lambda)$ is strictly increasing on $w \in (0,1-\lambda)$ and decreasing on $w \in (1-\lambda,2(1-\lambda))$. This with the boundary properties of $s(w,\lambda)$ implies that the national saving rate is hump-shaped on $w \in (0,2(1-\lambda))$ achieving its maximum at $w=1-\lambda$.

(b) It follows from the definition of the national saving rate that
\begin{equation}\label{eq_500_lemma_s_aggregate}
\pi(w,\lambda)=\frac{w}{w+2\lambda \phi(w,\lambda)}=\frac{1}{1+\frac{2\lambda \phi(w,\lambda)}{w}}.
\end{equation}
Monotonicity of $w \mapsto \frac{\phi(w,\lambda)}{w}$ implies that $w \mapsto \pi(w,\lambda)$ is a strictly increasing function. In addition
\begin{equation}\label{eq_600_lemma_s_aggregate}
\pi_1(w,\lambda)=s(w,\lambda)\left(1-\ds\frac{s(w,\lambda)w}{\psi(w,\lambda)}\left(\ds\frac{w}{1-\lambda}-1\right)\right).
\end{equation}
This with the boundary properties of $s(w,\lambda)$ implies the boundary properties of $\pi(w,\lambda)$.

(c) Monotonicity of $\lambda \mapsto s(w,\lambda)$ and $\lambda \mapsto \pi(w,\lambda)$ follows from Lemma \ref{lemma_theta} and from definitions of $s$ and $\pi$. \qed

\section{Time Discount and Flexible Investment Size\label{sec_robustness}}

\begin{figure}[ht!]
   \centering
   \subfigure[$s\left(\frac{w_t}{I},\lambda\right)$ ]
   {
   \includegraphics[width=0.4\textwidth]{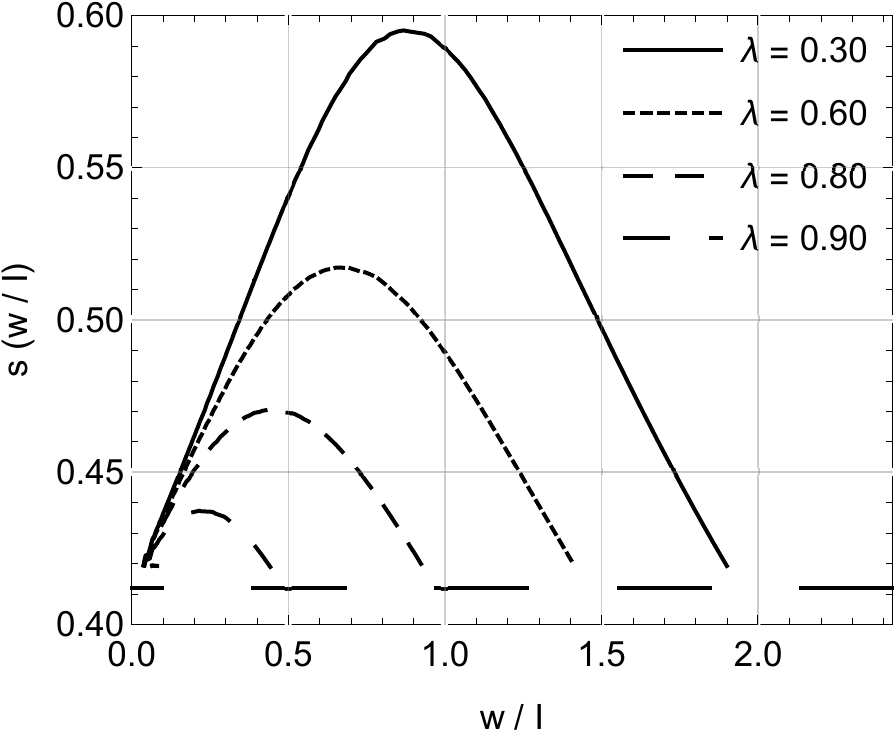}
   }
   \hspace{0.10cm}
   \subfigure[$\frac{1}{I}\pi\left(\frac{w_t}{I},\lambda\right)=s\left(\frac{w_t}{I},\lambda\right)\frac{w_t}{I}$]
   {
   \includegraphics[width=0.4\textwidth]{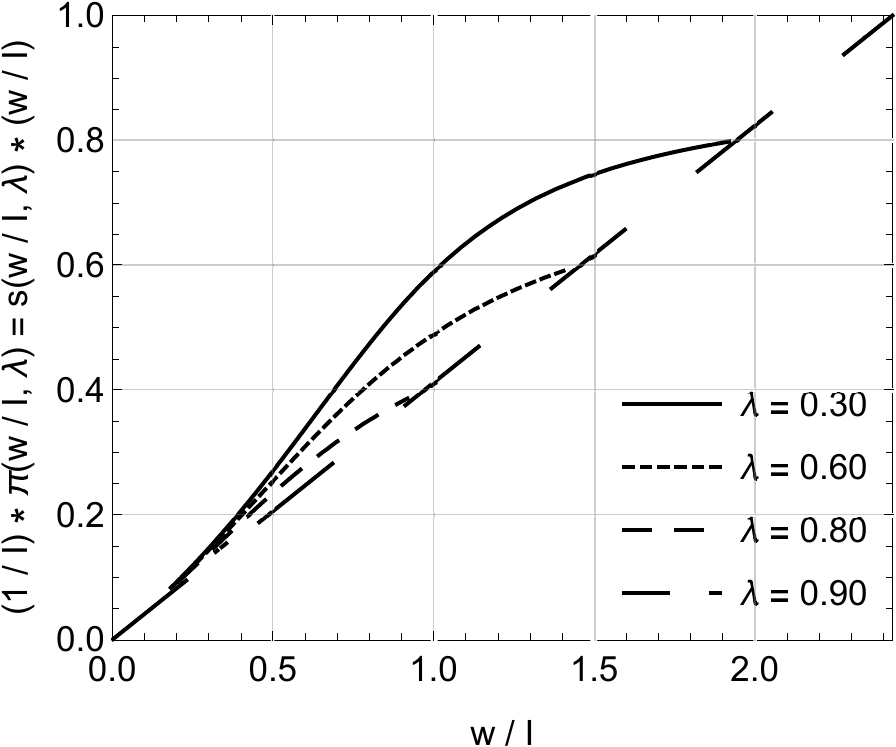}
   }
   \caption{The national saving rate and the fraction of entrepreneurs when $\beta=0.70$.}\label{fig:s-fraction-beta}
\end{figure}

This section shows that we can relax our assumptions of a zero time discount and a fixed investment size and still obtain essentially the same results. Suppose the agent's lifetime utility is $\ln c_{1t}+\beta \ln c_{2t+1}$ and that capital is produced by the following technology
\begin{equation*}
  F(i_t)=
\begin{cases}
  0 & \mbox{if $i_t<I$ }  \\
  Ri_t & \mbox{if $i_t\geq I$}
\end{cases}
\end{equation*}
where $i_t$ is the investment of the final good, $F(i_t)$ is the produced amount of capital, and $I$ is the minimum investment size. 
The lifetime utility of investors is $\ln U^{\ell}+\ln(w_t^{1+\beta}r_{t+1}^{\beta})$ where $U^{\ell}=\max_{s \in [0,1]}\{(1-s)s^{\beta}\}$. This implies that $s^{\ell}=\frac{\beta}{1+\beta}$ and $U^{\ell}=\frac{\beta^{\beta}}{(1+\beta)^{1+\beta}}$.
The lifetime utility of entrepreneurs is $\ln U^{b}(w_t/I,\phi_{t+1})+\ln(w_t^{1+\beta}r_{t+1}^{\beta})$ where
\begin{equation*}
U^{b}(w_t/I,\phi_{t+1},\lambda)=\max_{s \in [0,1]}\left\{\left(1-s\right)\left(\frac{\phi_{t+1}-1}{w_t/I}+s\right)^{\beta}\middle|\ s \geq \frac{1-\lambda \phi_{t+1}}{w_t/I}\right\}.
\end{equation*}
This implies that the optimal saving rate of entrepreneurs is
\begin{equation*}
  s^{b}_t=\max\left\{\frac{1}{1+\beta}\left(\beta-\frac{\phi_{t+1}-1}{w_t/I}\right),\frac{1-\lambda \phi_{t+1}}{w_t/I}\right\}
\end{equation*}
and
\begin{equation*}
U^{b}(w/I,\phi,\lambda)=\begin{cases}
\frac{\beta^{\beta}}{(1+\beta)^{1+\beta}}\left(1+\frac{\phi-1}{w/I}\right)^{1+\beta} & \textrm{if $\frac{w}{I} \geq 1-\frac{(1+\beta)\lambda-1}{\beta}\phi$}  \\[1.00ex]
\left(1-\frac{1-\lambda \phi}{w/I}\right)\left(\frac{(1-\lambda) \phi}{w/I}\right)^{\beta} & \textrm{if $\frac{w}{I} \in \left[1-\lambda\phi,1-\frac{(1+\beta)\lambda-1}{\beta}\phi\right)$}.
\end{cases}
\end{equation*}
If $\frac{w_t}{I}<1-\lambda\phi_{t+1}$, then young agents cannot become an entrepreneur because they cannot overcome the credit constraint even if they save the entire wage.
The equilibrium entrepreneurial rent is $\phi_{t+1}=1$ when $\frac{w_t}{I} \geq \frac{(1+\beta)(1-\lambda)}{\beta}$ and $\phi_{t+1}=\phi(\frac{w_t}{I},\lambda)$, which solves
\begin{equation*}
  \left(1-\ds\frac{1-\lambda \phi_{t+1}}{w_t/I}\right)\left(\ds\frac{(1-\lambda) \phi_{t+1}}{w_t/I}\right)^{\beta}=\ds\frac{\beta^{\beta}}{(1+\beta)^{1+\beta}}
\end{equation*}
when $1-\lambda\phi_{t+1}\leq\frac{w_t}{I}<\frac{(1+\beta)(1-\lambda)}{\beta}$. There is no closed form solution of the above equation when $\beta \neq 1$. However, we can show that the properties of $\phi$ demonstrated in Lemma \ref{lemma_theta} hold for $\beta \neq 1$ as well. 
The credit market clears when
\begin{equation*}
\frac{s_t w_t}{I}\left(I-s_t^{b}w_t\right)=\left(1-\frac{s_t w_t}{I}\right)\frac{\beta w_t}{1+\beta}.
\end{equation*}
The saving rate of entrepreneurs is
\begin{equation*}
s^{b}\left(\frac{w_t}{I},\lambda\right)=
\begin{cases}
  \frac{1-\lambda \phi(w_t/I,\lambda)}{w_t/I} & \textrm{if  $\frac{w_t}{I}<\frac{(1+\beta)(1-\lambda)}{\beta}$}      \\[1.00ex]
  \frac{\beta}{1+\beta}                         & \textrm{if $\frac{w_t}{I}\geq \frac{(1+\beta)(1-\lambda)}{\beta}$}.
\end{cases}
\end{equation*}
The fraction of entrepreneurs is
\begin{equation*}
s\left(\frac{w_t}{I},\lambda\right)=\begin{cases}
  \frac{\beta}{\beta \frac{w_t}{I}+(1+\beta)\lambda \phi(w_t/I,\lambda)}  & \textrm{if $ \frac{w_t}{I}<\frac{(1+\beta)(1-\lambda)}{\beta}$}      \\[1.00ex]
  \frac{\beta}{1+\beta}                               & \textrm{if $\frac{w_t}{I}\geq \frac{(1+\beta)(1-\lambda)}{\beta}$}.
\end{cases}
\end{equation*}
Figure \ref{fig:s-fraction-beta} shows the national saving rate and the fraction of entrepreneurs when $\beta=0.70$. The figure indicates that the properties of the saving rate hold under a more general specification of the basic model.

%\clearpage

%

%\clearpage

\section{Tables\label{sec:appendix-table}}
\begin{table}[ht!]
\scriptsize
\centering
\begin{tabular}{|l|l|l|l|l|l|}
        %{p{1.4cm}p{1.4cm}p{1.4cm}p{1.4cm}p{1.4cm}p{1.4cm}p{1.4cm}p{1.4cm}p{1.4cm}}
        \hline
Country & GDP p.c. ($y$)  & Credit/GDP ($\lambda$) & Country  & GDP p.c. ($y$) & Credit/GDP ($\lambda$) \\

        &[in Thousands] &                      &          & [in Thousands]&\\ \hline

Albania & 3.171 & 0.092 &Indonesia &3.142 &0.319\\ \hline
Algeria &2.912& 0.333  &Iran& 4.22 &0.227\\ \hline
Angola &3.053& 0.049  &Ireland & 13.204 & 0.604\\ \hline
Argentina &6.986& 0.185 &   Israel & 10.699 & 0.566\\ \hline
Armenia& 5.062 &0.079 & Italy& 13.643& 0.644 \\ \hline
Australia &14.364 &0.501 &Jamaica &4.439 &0.238\\ \hline
Austria &14.83& 0.738 &Japan& 14.311 &1.517\\ \hline
Azerbaijan &4.316& 0.064 &Mauritius &9.942& 0.428 \\ \hline
Bahrain &14.837& 0.442&Mexico &5.268 &0.224  \\ \hline
Bangladesh& 1.268& 0.175 & Mongolia& 1.87& 0.154 \\ \hline
Barbados& 13.165 &0.496 & Morocco& 2.468 &0.261  \\ \hline
Belarus &13.087 &0.127 & Mozambique &1.299 &0.131\\ \hline
Belize &5.672& 0.363 &Nepal &0.897& 0.129\\ \hline
Benin& 0.746& 0.156 &Netherlands &14.625& 0.863\\ \hline
Bolivia& 1.915 &0.258&New Zealand& 11.354& 0.552\\ \hline
Bosnia \& Herz. &4.445 &0.426     & Nicaragua& 1.601& 0.25              \\ \hline
Brazil &4.604 &0.423  & Niger & 0.588 &0.094 \\ \hline
Bulgaria & 6.262& 0.368&Nigeria & 0.849 &0.11 \\ \hline
Burkina Faso& 0.663& 0.108&Norway &16.985& 0.484\\ \hline
Burundi& 0.506 &0.097 & Oman& 10.919& 0.249\\ \hline
Cambodia& 1.819 &0.074&Pakistan& 1.499 &0.245\\ \hline
Cameroon &1.491 &0.167&Panama &3.426& 0.606 \\ \hline
Canada &15.04 &0.817&Papua New Guinea &1.459 &0.185\\ \hline
Central Afr. Rep.& 0.651& 0.104&Paraguay &2.791& 0.198\\ \hline
Chad &0.874 &0.079   &Peru& 2.984& 0.168 \\ \hline
Chile &6.164& 0.442 & Philippines& 2.084& 0.271\\ \hline
China &2.624 &0.874 &Poland& 8.417 &0.276\\ \hline
Colombia& 3.42& 0.283   &Portugal& 8.506& 0.749\\ \hline
Congo, Rep. of &1.421 &0.145 &Qatar& 30.938& 0.299\\ \hline
Costa Rica &5.111& 0.234 &Romania& 6.685 &0.149\\ \hline
Croatia &9.394& 0.401  &Russia &8.37& 0.187 \\ \hline
Cyprus &12.216 &1.238    & Rwanda& 0.794& 0.063 \\ \hline
Czech Republic& 15.303 &0.487 & Samoa &3.888 &0.243\\ \hline
Denmark &14.335 &0.642  &Senegal &1.144 &0.22\\ \hline
Djibouti &3.651 &0.354    &Sierra Leone &1.326& 0.048\\ \hline
Dom. Republic &3.527& 0.231&Singapore& 14.802& 0.744 \\ \hline
Ecuador &3.029 &0.217 &Slovenia& 16.958 &0.382 \\ \hline
Egypt &2.282 &0.28&South Africa& 5.125 &0.902 \\ \hline
El Salvador &2.88& 0.303 & Spain & 11.61 & 0.79 \\ \hline
Eq. Guinea &5.549 &0.096 &Tajikistan& 2.2 & 0.173\\ \hline
Estonia & 11.733 & 0.494  &Tanzania& 0.626 &0.087   \\ \hline
Ethiopia &0.746 &0.151  & Thailand &3.498 &0.657\\ \hline
Fiji &2.983 &0.286 &     Togo &0.684& 0.185\\ \hline
Finland &13.084& 0.565 &Trinidad \&Tobago& 7.552& 0.336\\ \hline
France& 13.184 &0.806& Tunisia& 3.919& 0.514 \\ \hline
Gabon& 4.584 &0.147  & Turkey &3.179& 0.182\\ \hline
Gambia, The& 0.823 & 0.134 &Turkmenistan &6.589 &0.017\\ \hline
Georgia &4.983 &0.105& Uganda &0.58 &0.066 \\ \hline
Germany &17.353 &0.913 & Ukraine& 6.208 &0.191\\ \hline
Ghana& 0.925 &0.071& United Arab Emir. &32.473& 0.29 \\ \hline
Greece& 10.546& 0.354& United Kingdom& 13.117 &0.737\\ \hline
Guatemala& 3.043& 0.167& United States &18.805 &1.219\\ \hline
Guinea-Bissau &0.641& 0.088  & Uruguay &5.732& 0.323\\ \hline
Guyana &1.562 &0.333& Venezuela &5.365 &0.29 \\ \hline
Haiti &1.431& 0.138& Vietnam &2.432& 0.43\\ \hline
Honduras& 1.898 &0.294 & Yemen& 0.928 &0.055\\ \hline
Hungary& 10.36 &0.399& Zambia &1.006 &0.119\\ \hline
Iceland &15.676& 0.628 & Zimbabwe &2.238& 0.298\\ \hline
India &1.307 &0.213  &&&\\ \hline

\end{tabular}
\caption{\label{tab:appendix-1} List of countries }
\end{table}

\section{Robustness\label{sec:robustness}}  

In Table \ref{tab:3}, we examine the sensitivity of our IV estimates
to excluding from the sample large oil importing and exporting countries
(column (1)); excluding the top and bottom 1st percentile of the change
in the national saving rate (column (2)); using initial (1970) oil
net-export GDP shares to construct the oil price shock instrument
(column (3)); and splitting the sample into the post-1990 and pre-1990
period (columns (4) and (5)).\footnote{Table \ref{tab:appendix-2} 
	shows first stage effects of the oil price instrument on the change
	in GDP per capita across these robustness checks.} The main result from these robustness checks is that the estimated
coefficient of $\theta$ ($\gamma$) is negative (positive) and significantly
different from zero at the conventional significance levels.

\begin{table}[ht!]
	\singlespacing
	\footnotesize
	\begin{center}
		\begin{tabular}{lllllll}
			\hline
			
			\multicolumn{6}{c}{$\Delta\ln(s_{it})$} \\
			
			\hline
			& (1) & (2) & (3) &  (4) & (5) \\
			
			& Excluding large  & Excluding  &  Using initial oil& Pre-1990 & Post-1990 \\
			
			& importers \& exporters & top/botton 1st pctl. &  net-export shares&  & \\

			$\Delta\ln(y_{it})$    & 4.40***    &2.31***&  2.27*** & 2.04**  & 3.23*** \\
			
			& (1.44)   & (0.57) & (0.71) & (0.98)   & (0.83) \\

			$\Delta\ln(\lambda_{it})$    & -0.20*   &-0.17**&  -0.16** & -0.43*  & -0.25** \\
			
			& (0.11) & (0.09) & (0.07) &  (0.22)  & (0.10)  \\

			Kleibergen-Paap   &  13.10    & 16.80     & 17.87& 9.41 & 169.33 \\
			
			F-stat  & &&&&\\

			Cragg-Donald   &  70.11    & 282.15     & 268.30& 134.12 & 261.13 \\
			
			F-stat   &&&&\\
			
			Endogenous     &   $\Delta\ln(y_{it})$  &   $\Delta\ln(y_{it})$ &  $\Delta\ln(y_{it})$ & $\Delta\ln(y_{it})$ & $\Delta\ln(y_{it})$ \\
			
			Regressors   &   &&&&\\

			Instruments & $OPS_{it}$ & $OPS_{it}$  & $OPS_{it}$  &$OPS_{it}$  & $OPS_{it}$ \\
			
			Country FE  &  Yes    & Yes   & Yes&Yes &  Yes     \\

			Year FE        & Yes  & Yes  & Yes & Yes& Yes    \\
			
			Observations               & 3034  & 3362 & 3721 &2075&1622   \\
			
			\hline
			
		\end{tabular}
	\end{center}
	
	Note: The dependent variable, $\Delta\ln(s_{it})$, is the change in the log of the savings rate. $\Delta\ln(y_{it})$ is the change in the log of real GDP per capita; $\Delta\ln(\lambda_{it})$ is the change in the log of the credit-to-GDP ratio. The method of estimation is two-stage least squares. Huber robust standard errors (shown in parentheses) are clustered at the country level. Column (1) excludes large oil importing countries (China, France, Italy, Japan, South Korea, Netherlands, United Kingdom, and United States) and large oil exporting countries (Algeria, Canada, Indonesia, Iran, Iraq, Kuwait, Libya, Mexico, Nigeria, Norway, Oman, Qatar, Russia, United Arab Emirates, and Venezuela). Column (2) excludes observations in the top and bottom 1st Percentile of $\Delta\ln(s_{it})$. Column (3) uses 1970 oil net-export shares to construct the oil price shock instrument. Column (4) shows estimates for the pre-1990 period; column (5) post-1990 period.
	\vspace{1em}
	\caption{\label{tab:3} Effects of growth in GDP per capita and the credit-to-GDP ratio on the growth rate of the savings rate (robustness to excluding outliers; excluding large oil importers and exporters; using initial oil net-export GDP shares; time-period split)
	}
\end{table}

\begin{table}%[ht!]
	\singlespacing
	\scriptsize
	
	\begin{center}
		\begin{tabular}{lcccccccc}
			\hline
			
			\multicolumn{9}{c}{$\Delta\ln(y_{it})$} \\
			
			\hline
			& (1)        & (2)      & (3)      &  (4)    & (5) & (6)& (7) &(8)\\
			
			& Full Sample        & Excluding      & Excluding        &  Using      & Pre-  & Post- & Rich  & Poor\\
			
			&        &  Large       &  Top/       &   Initial Oil     & 1990  & 1990 &  Countries &  Countries\\
			
			&        & Importers \&       & Bottom       &  Net-Export    & Period& Period& & \\
			
			& &  Exporters      &   1st Pctl.    &    Shares   &  &  &  & \\

			$OPS_{it}$                              & 1.01***    & 0.98***  & 0.98*** & 6.24***  & 0.79*** &2.45*** &1.01*** & 1.36***\\
			
			& (0.23)&(0.27)&(0.23)&(1.52)&(0.26)&(0.19)&(0.25)& (0.35)\\

			Country FE                              &  Yes       & Yes       & Yes    & Yes              & Yes  & Yes             & Yes & Yes \\
			
			Year FE                                 & Yes        & Yes       & Yes    & Yes              & Yes  & Yes        & Yes & Yes      \\
			
			Observations                                       & 3781       & 3034     & 3721  & 3362             &2075 & 1622 &1890&1891\\
			
			\hline
		\end{tabular}
	\end{center}
	
	Note: The dependent variable, $\Delta\ln(y_{it})$,
	is the change in the log of real GDP per capita. The method of estimation
	is least squares. Huber robust standard errors (shown in parentheses)
	are clustered at the country level. Column (2) excludes large oil
	importing countries (China, France, Italy, Japan, South Korea, Netherlands,
	United Kingdom, and United States) and large oil exporting countries
	(Algeria, Canada, Indonesia, Iran, Iraq, Kuwait, Libya, Mexico, Nigeria,
	Norway, Oman, Qatar, Russia, United Arab Emirates, and Venezuela).
	Column (3) excludes observations in the top and bottom 1st Percentile
	of $\Delta\ln(s_{it})$. Column (4) uses 1970 oil net-export shares
	to construct the oil price shock instrument. Column (5) shows estimates
	for the pre-1990 period; column (6) post-1990 period. Column (7) shows
	estimates for countries with above sample median GDP per capita; column
	(8) below sample median GDP per capita.
	
	\vspace{1em}
	\caption{\label{tab:appendix-2} First stage effects}
\end{table}

\clearpage

\section{Post Global Financial Crisis\label{sec:post-global-crisis} }

Column (1) of Table \ref{tab:2007-2017} shows estimates of the model
over the period 2007-2017, which starts with the
global financial crisis. The data are from the World Development Indicators \citep{bank2020world}. For comparison, column (2) of Table \ref{tab:2007-2017}
shows estimates for the period that excludes the global financial
crisis, i.e. from 1960 to 2006. The estimates in Table \ref{tab:2007-2017}
are generated from least squares regressions. There are two reasons
why we report least squares estimates. First, least squares are more
efficient for testing whether the coefficients of interest are the
same in the two periods. Second, instrumental variables regressions
(not reported) do not yield a significant first stage relationship
over the period 2007-2017.

One can see from Table \ref{tab:2007-2017} that the estimated coefficients
in columns (1) and (2) are qualitatively similar. The hypothesis that
the coefficients in column (1) are the same as in column (2) cannot
be rejected at the conventional significance levels. Hence, estimation
of the model for the time period that includes the financial crisis
yields similar results as least squares estimation of the model that
excludes this time period.

\begin{table}[ht!]
	\singlespacing
	\footnotesize
	
	\begin{center}
		\begin{tabular}{lll}
			\hline
			
			\multicolumn{3}{c}{$\Delta\ln(s_{it})$} \\
			
			\hline
			& (1)        & (2)     \\
			
			Time Period & 2007-2017& 1960-2006 \\
			
			$\Delta\ln(y_{it}) $    & 3.364***    & 1.941*** \\
			
			& (0.708)&(0.233)\\

			$\Delta\ln(y_{it})*y_i$   & -0.028***   & -0.045*** \\
			
			& (0.011)& (0.011)\\
			
			$\Delta\ln(y_{it})*\lambda_i$     & -1.935***  & -2.982 \\
			
			& (0.74)&(4.31)\\
			
			$\Delta\ln(\lambda_{it})$        &  -0.707***   & -0.309*** \\
			
			& (0.159) &(0.111)\\

			Country FE         &  Yes       & Yes                  \\
			
			Year FE             & Yes        & Yes                 \\
			
			Observations       & 887       & 3787   \\
			
			\hline
		\end{tabular}
	\end{center}
	
	Note:  The method of estimation is least squares. Huber robust standard errors (shown in parentheses) are clustered at the country level. The dependent variable is the change in the log of the national savings rate. *Significantly different from zero at the 10 percent significance level, ** 5 percent significance level, *** 1 percent significance level. \vspace{1em}
	\caption{\label{tab:2007-2017} Estimates for post-crisis period: 2007-2017
	}
\end{table}

\section{Clustering\label{sec:clustering}}

We have carried out an additional sensitivity analysis where we clustered countries according to their GDP per capita, geographic location, population size, and sectoral composition. We then estimated the econometric model on the clusters. Since we want to obtain estimates that reflect a plausibly causal effect of economic growth on the saving rate, we use instrumental variables methods. Specifically, the method of estimation is two-stage least squares. In the 2SLS estimation, GDP per capita growth (and the interaction with average GDP per capita) is instrumented by the growth rate of a country-specific international commodity net-export price index (and the interaction of the latter variable with average GDP per capita). The commodity price index index is constructed as the geometric average of the international commodity prices. The geometric weights are countries' average GDP shares of net exports of the commodities. The index is thus country-specific: the larger the net-export GDP share of a commodity the larger is the effect of growth in the international commodity price. The identifying assumption is that economic growth of a country does not affect international commodity prices. This assumption is satisfied for most countries: individually, each country's demand or supply of a commodity is small relative to the world market; it is thus reasonable to assume that the majority of countries are price takers on the international commodity market.

We have estimated the econometric models on the largest sample. This sample is determined by the available data from the Penn World Tables (for real GDP per capita at chained PPPs and the national saving rate) and IMF commodity statistics (for the commodity net-export price index). The sample spans the period 1960 to 2019 and covers 153 countries.

One issue that arises with the clustering sensitivity analysis is that, for each cluster, there is a smaller number of observations relative to the whole sample. To balance the trade-off between parameter heterogeneity and the smaller sample size for each cluster, we estimate a more parsimonious econometric model that only includes on the right-hand side GDP per capita growth and the interaction with countries' average GDP per capita (controlling for country and time fixed effects). This model enables us to examine whether, for each cluster, the effect of economic growth on the saving rate depends on countries' average income.

\clearpage

Table \ref{table:sensitivity} shows results when countries are clustered by GDP per capita, geographic location, population size, advanced economy status, and sectoral composition. For comparison, we report in columns (1) and (2) of Table \ref{table:sensitivity} estimates of the econometric model for the largest sample, i.e. without grouping countries into clusters. The clustering for the remaining columns in Table \ref{table:sensitivity} is then as follows:
\begin{itemize}
	\item above vs. below median GDP per capita, see columns (3) and (4);
	\item 	countries which form part of the Middle-East-and-Africa region vs. countries that are not part of that region, see columns (5) and (6);
	\item 	above vs. below median population size, see columns (7) and (8);
	\item 	advanced economies vs. emerging market and low income economies, see columns (9) and (10)
	\item 	countries with a GDP share of agriculture less than 20\% or 10\%, see columns (11) and (12)
\end{itemize}

In column (13) of Table \ref{table:sensitivity}, we report estimates for the whole sample when controlling for sectoral composition. The variable that measures sectoral composition is the GDP share of agriculture (obtained from the World Development Indicators).

For all columns of Table \ref{table:sensitivity}, one can see that the estimated coefficient on GDP per capita growth is positive. The coefficient on the interaction between GDP per capita growth and countries' average GDP per capita during 1960 to 2019 is negative. The interpretation of these estimates is that the effect of GDP per capita growth on the saving rate is positive, and decreasing in countries' average GDP per capita: only in countries with relatively low average GDP per capita does economic growth have a significant positive effect on the saving rate.

The instruments are relevant as judged by the Kleibergen Paap or Cragg Donald F-statistics: In most columns, these F-stats are above 10. One exception was for advanced economies: for this group, the first stage F-stats are very low -- 0.5 for the Kleibergen Paap F-stat and 1.3 for the Cragg Donald F-stat. Given these very low F-stats we reported in column (9) of Table \ref{table:sensitivity} least squares estimates and not two-stage least squares estimates

\begin{table}[ht!]
	{
		\center
		{
			\scalebox{.7}{\includegraphics[angle=-90]{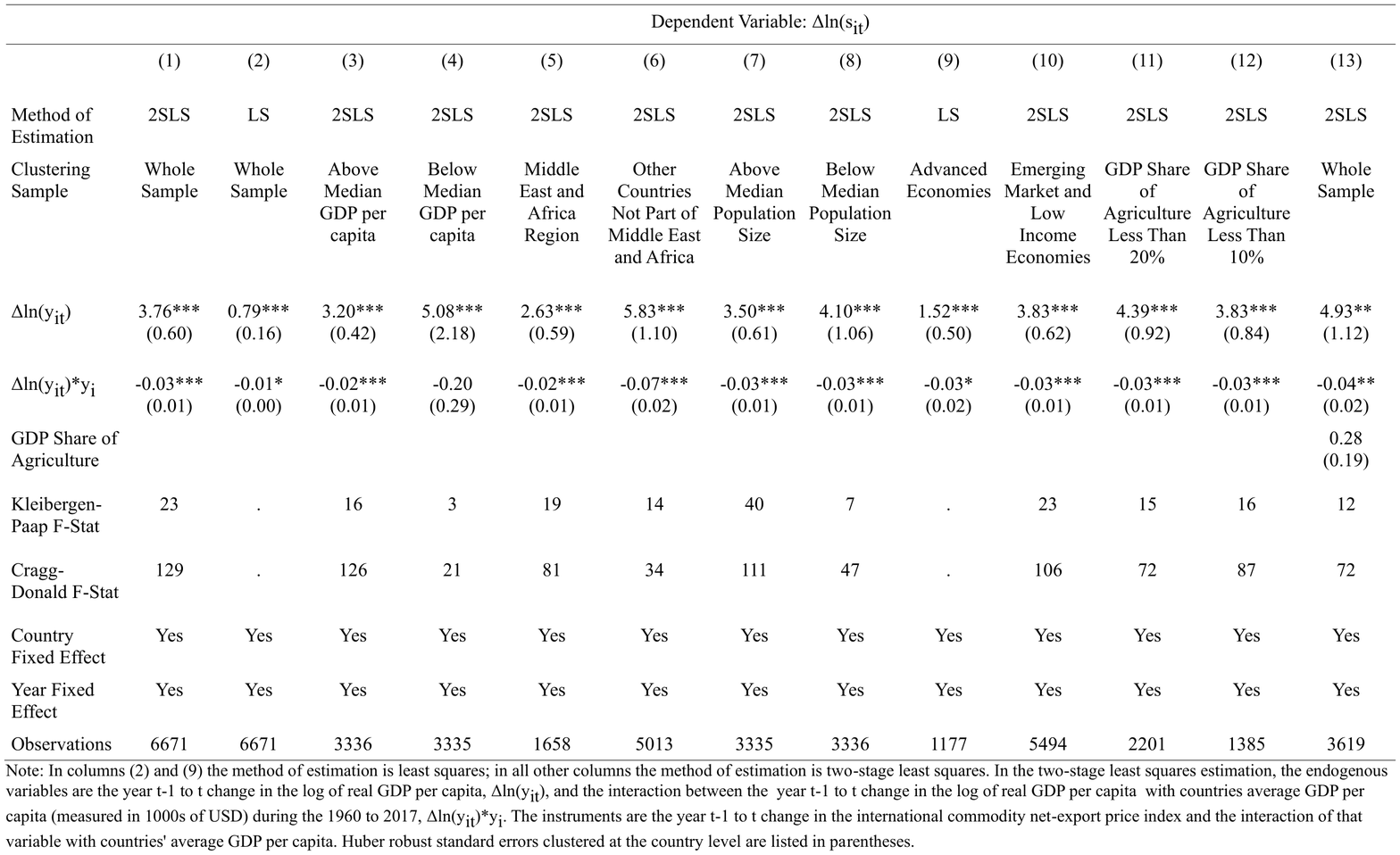}}
		}
		
		\caption{Effects of Economic Growth on the Domestic Saving Rate} \label{table:sensitivity}
	}
	
\end{table}

\clearpage{}

\bibliographystyle{ecta} 
\bibliography{localbib}

\end{document}